\documentclass[10pt,twocolumn,twoside,journal]{IEEEtran}
\usepackage[T1]{fontenc}
\pdfoutput=1
\usepackage[cmintegrals]{newtxmath}

\usepackage{cite}
\usepackage{algorithmic}

\usepackage{algorithm} 
\makeatletter
\newcommand{\removelatexerror}{\let\@latex@error\@gobble}
\makeatother

\usepackage{graphicx}
\usepackage{textcomp}
\usepackage{xcolor}

\usepackage{lipsum}     
\usepackage{multicol}   

\usepackage{booktabs}
\newcommand\tstrut{\rule{0pt}{2.4ex}}
\newcommand\bstrut{\rule[-1.0ex]{0pt}{0pt}}
\usepackage{balance}

\ifCLASSOPTIONcompsoc
\usepackage[caption=false,font=normalsize,labelfon
t=sf,textfont=sf]{subfig}
\else
\usepackage[caption=false,font=footnotesize]{subfig}

\usepackage{multirow}
\usepackage{array}

\begin{document}
\title{Efficient Joint DOA and TOA Estimation for Indoor Positioning with 5G Picocell Base Stations}

\author{Mengguan~Pan,
  Peng~Liu,
  Shengheng~Liu,~\IEEEmembership{Senior Member,~IEEE,}
  Wangdong~Qi,~\IEEEmembership{Member,~IEEE,}
  Yongming~Huang,~\IEEEmembership{Senior Member,~IEEE,}
  Xiaohu~You,~\IEEEmembership{Fellow,~IEEE,}
  Xinghua~Jia,
  Xiaodong~Li
  \vspace{-1.6em}
  \thanks{Manuscript received XXX XX, 2022. This research was supported in part by the National Natural Science Foundation of China under Grant Nos. 62001103 and U1936201.}
        \thanks{The authors are with the Purple Mountain Laboratories, Nanjing 211111, China. (Emails: panmengguan@pmlabs.com.cn, herolp@gmail.com, s.liu@seu.edu.cn, qiwangdong@pmlabs.com.cn, huangym@seu.edu.cn, xhyu@seu.edu.cn)}
        \thanks{P.~Liu is also with the Nanjing University of Aeronautics and Astronautics, and the Army Engineering University of PLA, Nanjing 210007, China.}
        \thanks{S.~Liu, W.~Qi, Y.~Huang, and X.~You are also with the National Mobile Communications Research Laboratory, Southeast University, Nanjing 210096, China.}}

\markboth{IEEE TRANSACTIONS ON INSTRUMENTATION AND MEASUREMENT, ~Vol.~XX, No.~X, XXX~2022}%
{PAN \MakeLowercase{\textit{et al.}}: Efficient Joint DOA and TOA Estimation for Indoor Positioning with 5G Picocell Base Stations}

\maketitle
\newcommand{\figpath}{./figs/}

\begin{abstract}
  The ubiquity, large bandwidth, and spatial diversity of the fifth generation (5G) cellular signal render it a promising candidate for accurate positioning in indoor environments where the global navigation satellite system (GNSS) signal is absent. In this paper, a joint angle and delay estimation (JADE) scheme is designed for 5G picocell base stations (gNBs) which addresses two crucial issues to make it both effective and efficient in realistic indoor environments.
  Firstly, the direction-dependence of the array modeling error for picocell gNB as well as its impact on JADE is revealed. This error is mitigated by fitting the array response measurements to a vector-valued function and pre-calibrating the ideal steering-vector with the fitted function.
  Secondly, based on the deployment reality that 5G picocell gNBs only have a small-scale antenna array but have a large signal bandwidth,
  the proposed scheme decouples the estimation of time-of-arrival (TOA) and direction-of-arrival (DOA) to reduce the huge complexity induced by two-dimensional joint processing. It employs the iterative-adaptive-approach (IAA) to resolve multipath signals in the TOA domain, followed by a conventional beamformer (CBF) to retrieve the desired line-of-sight DOA.
  By further exploiting a dimension-reducing pre-processing module and accelerating spectrum computing by fast Fourier transforms, an efficient implementation is achieved for real-time JADE.
  Numerical simulations demonstrate the superiority of the proposed method in terms of DOA estimation accuracy. Field tests show that a triangulation positioning error of 0.44 m is achieved for 90\% cases using only DOAs estimated at two separated receiving points.
\end{abstract}


\IEEEpeerreviewmaketitle

\section{Introduction}
\IEEEPARstart{L}{ocation} awareness plays a paramount role in a wealth of scenarios, such as autonomous driving \cite{kuutti2018_SurveyStateoftheArt}, intelligent transportation \cite{ding2022_SmartLOCIndoorLoca}, emergency relief \cite{goncalvesferreira2017_LocalizationPositio}, assisted living \cite{witrisal2016_HighAccuracyLocaliz}, etc., in the era of Internet-of-everything (IoE). In outdoor environments, the global navigation satellite systems (GNSS) provide robust and accurate positioning information, while in deep urban canyons and indoor environments, they are unreliable owing to the severe blockage of the line-of-sight (LOS) signals.

Recently, considerable attention has been devoted to employ other radio frequency (RF) signals, such as ultra-wideband (UWB) signals \cite{flueratoru2021_HighAccuracyRanging, jimenezruiz2017_ComparingUbisenseB}, Bluetooth signals \cite{ho2021_BlueTrkTrackingPrea, yang2022_BluetoothIndoorLoca}, Wi-Fi signals \cite{he2016_WiFiFingerprintBase, banin2019_ScalableWiFiClient}, radio frequency identification (RFID) signals \cite{li2019_ReviewUHFRFID, li2021_ReLocUHFRFIDRelati}, or signals from cellular networks \cite{delperal-rosado22_SurveyCellularMobi, he2021_MultiBSSpatialSpec}, for high-accuracy indoor positioning. Among them, the fifth-generation new radio (5G NR) signal is particularly noteworthy from the following aspects. First, since the 5G picocell base stations (a.k.a gNodeBs, or gNBs) are being, or will be densely deployed for indoor coverage, 5G signals will be abundant in indoor scenarios. Second, the large bandwidth and high received power of 5G signals and the widely employed antenna array technology for 5G transmission/reception points (TRPs) benefit the localization parameter estimation especially \cite{kassas2017_HearThereforeKnow}. Lastly, although 5G networks are designed for the purpose of communications, there exists reference signals dedicated for positioning and sensing in the up-to-date 3GPP standard, i.e. the sounding reference signal (SRS) for positioning in the uplink and the positioning reference signal (PRS) for positioning in the downlink \cite{3gpp.38.857}. This renders it viable to implement integrated localization and communications (ILAC) by 5G infrastructure without modifying the underlying hardware and the upper-layer protocol \cite{xiao2020_OverviewIntegrated}.

This paper focuses on the joint estimation of the direction-of-arrival (DOA) and the time-of-arrival (TOA), which is also known as the joint angle and delay estimation (JADE), based on 5G picocell gNBs for the purpose of indoor positioning. The main challenges are twofold. First, the small-scale antenna arrays established for picocell gNBs suffer from much lower spatial resolution and much severer array modeling errors than large-scale antenna arrays, both of which limit the DOA estimation accuracy. Second, the parameter space for JADE is huge and the dimension for the vectorized space-time or space-frequency data is high, which make JADE algorithms unfavorable for real-time implementations. Neither of these challenges has been fully addressed in previous work. 

\subsection{Related Works}
The JADE problem was first addressed in \cite{vanderveen1997_JointAngleDelay} for the global system for mobile communications (GSM) signal, also known as the second-generation (2G) cellular signal, in which the two-dimensional (2-D) multiple signal classification (MUSIC) and the maximum likelihood (ML) method are utilized to estimate the DOA and TOA jointly. After that, the MUSIC-based solution becomes popular for JADE problems in different scenarios \cite{kotaru2015_SpotFiDecimeterLeva, bazzi2016_SpatiofrequentialSm, sun2021_ComparativeStudy3D}. Since the MUSIC algorithm is originally derived on the assumption of non-coherent impinged signals, additional pre-processing techniques must be developed for coherent sources, which is the case for an indoor environment with dense multipath reflections. In particular, a similar JADE approach which combines the 2-D MUSIC algorithm with an elaborately designed spatial-frequential smoothing technique is proposed for Wi-Fi signals and general orthogonal frequency division multiplexing (OFDM) signals, respectively in \cite{kotaru2015_SpotFiDecimeterLeva} and \cite{bazzi2016_SpatiofrequentialSm}. However, the eigen value decomposition (EVD) of the high-dimensional covariance matrix and the 2-D spectrum search on the DOA-TOA plane performed by MUSIC-based JADE methods are both computational intensive.

To reduce the overhead, search-free subspace-based methods which utilize the shift-invariance properties of the received signal in spatial and temporal/frequential domains were proposed in the JADE literature, such as that based the estimation of signal parameters via rotational invariant technique (ESPRIT) \cite{vanderveen1997_JointAngleDelaya, hua2016_JointEstimationDOA} and that based on the matrix-pencil method \cite{li2021_DecimeterLevelIndo, bazzi2016_SingleSnapshotJoin, shamaei2021_JointTOADOAa}. Specifically, \cite{vanderveen1997_JointAngleDelaya} and \cite{hua2016_JointEstimationDOA} derive the ESPRIT-based JADE method for GSM signals and UWB signals, respectively; \cite{li2021_DecimeterLevelIndo}, \cite{bazzi2016_SingleSnapshotJoin}, and \cite{shamaei2021_JointTOADOAa} derive the matrix-pencil-based JADE method for general OFDM signals, Wi-Fi signals, and the fourth-generation (4G) long-term evolution (LTE) signals, respectively. When in the presence of coherent sources, ESPRIT-based JADE method must employ additional pre-processing technique to recover the rank of the covariance matrix, similar to the MUSIC-based JADE method, while the matrix-pencil-based one can be directly applied to the original data. Although these search-free subspace-based JADE methods obviate the 2-D spectrum search and achieve a much lower complexity than the MUSIC-based solution, the EVD or the singular value decomposition (SVD) of a high-dimensional matrix is still needed.

Compared with the aforementioned subspace-based JADE methods, the ML-based JADE methods achieve a better statistical performance when the number of snapshots is small and can be directly applied for coherent sources \cite{stoica1989_MUSICMaximumLikeli}. However, they suffer from exhaustive computational burden as they require a \(pK\)-dimensional search, where \(K\) and \(p\) represent the number of impinged sources and the number of signal parameters to be estimated in each source. There have been several research efforts on developing computational attractive solutions for ML-based JADE methods \cite{wax1997_JointEstimationTim, kim2017_AOATOABasedLocaliz, qian2018_Widar2PassiveHuman, sun2021_ComparativeStudy3D, bazzi2015_EfficientMaximumLi}. The first category transforms the complicated high-dimensional ML search into several successive low-dimensional search based on the idea of alternating minimization \cite{wax1997_JointEstimationTim, kim2017_AOATOABasedLocaliz} or expectation-maximization (EM) \cite{qian2018_Widar2PassiveHuman, sun2021_ComparativeStudy3D}; while the second category derives a close-form solution for each ML iteration based on
polynomial parameterizations \cite{bazzi2015_EfficientMaximumLi}.
The most widely used ML implementation in the first category is the space-alternating generalized EM (SAGE) method \cite{fessler1994_SpacealternatingGen}, which has been employed in \cite{qian2018_Widar2PassiveHuman} and \cite{sun2021_ComparativeStudy3D} for JADE based on Wi-Fi signals and 5G signals, respectively. As an example for the second category, \cite{bazzi2015_EfficientMaximumLi} derives the 2-D iterative quadratic ML (IQML) algorithm \cite{bresler1986_ExactMaximumLikeli} for JADE based on OFDM signals. Other polynomial parameterization methods for ML estimators can also be extended to solve the JADE problem \cite{stoica1990_NovelEigenanalysis}. However, computation times for these algorithms can still be long owing to the large number of iterations before convergence.

There are also several efforts in modeling the JADE as a sparse inverse problem and seeking the solution according to the compressed sensing (CS) theory \cite{gong2017_RobustIndoorWirelea, zhang2019_BreathTrackTracking, fan2021_LocalizationBasedI, liu2021_IndoorPassiveLocala}. When certain conditions are met, it is possible for sparse recovery methods to recover DOAs and TOAs from fewer measurements than those needed by the subspace-based and ML-based methods. In addition, CS-based estimators do not require the knowledge of the number of signal sources. However, CS-based JADE solutions also exhibit high computational burdens since they have to deal with the huge-scale 2-D overcomplete dictionary and, depending on the underlying sparse recovery algorithms, they usually have to solve a large-dimensional convex optimization problem \cite{gong2017_RobustIndoorWirelea, zhang2019_BreathTrackTracking} or perform complicated matrix operation iteratively \cite{fan2021_LocalizationBasedI, liu2021_IndoorPassiveLocala}. 

These existing JADE algorithms all assume the perfect antenna response. However, the actual receiving signals from real-world antenna arrays are inevitably impaired by mutual coupling effects, sensor location perturbations, the gain-phase mismatch of the array multichannel receiver and other unpredictable effects \cite{ferreol2010_StatisticalAnalysis}. Deviations of the actual array manifold from the ideal one are referred as the array modeling errors and they can degrade the estimation performance severely \cite{wax2021_LocalizationMultipla}.
Further, the actual manifold cannot preserve the Vandermonde structure even for uniform arrays, which renders many aforementioned algorithms infeasible, such as the spatial-frequential smoothing technique and the ESPRIT-based, the matrix-pencil-based, and the IQML-based JADE algorithms.

\subsection{Contributions}
In this paper, an efficient JADE scheme for 5G picocell gNBs
is proposed to address the aforementioned two main challenges. First, the array modeling error is counteracted by employing the actual spatial steering-vector function for JADE, which is obtained by offline pre-calibrating the ideal steering-vector function using the array response measured in a multipath-free environment. Then based on the fact that picocell gNBs exhibit a much better TOA resolution than DOA resolution owing to its large signal bandwidth and small-scale antenna array, the proposed scheme achieves a largely reduced complexity by decoupling the subcarrier and space domains of the multichannel OFDM signal and performing TOA and DOA estimation cascadingly.
Specifically, a TOA spectral estimator based on the iterative adaptive approach (IAA) \cite{yardibi2010_SourceLocalization} is first applied to each receiving channel to segregate multipath components, followed by a conventional beamformer (CBF) to extract DOA information. After that, two enhancements to the proposed scheme are further presented to facilitate its real-time implementation. The first one is a pre-processing module which reduces the number of the frequency-domain samples based on the fact that the coverage of a picocell gNB is several orders lower than its unambiguous measurable range.
The latter is the accelerating of the IAA-based TOA spectral estimator by exploiting the fast Fourier transform (FFT) based on the evenly spaced subcarrier pattern of the 5G reference signal.


The technical contributions of this work are summarized below.
\begin{enumerate}
\item We investigate the array modeling errors of picocell gNBs based on real-measured data from commercial 5G equipment. To our best knowledge, this is the first work to reveal their direction-dependent characteristics and their impact on the JADE performance. 
\item We characterize the direction-dependent antenna errors as a vector-valued function, which fully captures all sorts of array modeling errors and is more expressive than the popular approach used in the DOA estimation literature which describes each error source separately by a limited number of parameters \cite{liu2016_SparseBasedApproach}. Based on this model, we also propose a calibration approach which pre-calibrates the steering-vector function at offline stage, and when this pre-calibrated function is used at online stage, the antenna errors are counteracted for signals from any direction.
\item We propose a JADE scheme whose real-time implementation is guaranteed. Its complexity has been reduced by three approaches: a cascading estimation scheme (Section \ref{sec:proposed_jade}), dimension reduction by pre-processing (Section \ref{sec:cfr-denoise}) and FFT accelerating (Section \ref{sec:fastIAA}). Its averaged running time is \(10.1\;\mathrm{ms}\) on a personal computer (PC) for typical picocell gNB configurations, which is nearly three orders lesser than that of the MUSIC-based JADE method.
\end{enumerate}

In addition, to evaluate the proposed JADE scheme, comprehensive experiments, including numerical simulations based on the signal model and based on 3GPP 5G indoor channel models \cite{3gpp.38.901} and field tests in an anechoic chamber and in a realistic indoor environment, are conducted. Experiments in a multipath-free anechoic chamber demonstrate that the proposed JADE scheme can substantially reduce the DOA bias caused by the direction-dependent antenna errors, especially in large incident directions. For example,
the averaged DOA estimation error can be reduced from \(8.11^\circ\) to \(1.28^\circ\) at an incident angle of \(+60^\circ\). According to numerical simulation results, when compared with existing 2-D super-resolution JADE methods, the proposed method has a significant improvement for the DOA estimation performance at a cost of a slightly reduced TOA estimation performance.
Field test in an indoor environment shows that in \(90\%\) cases, a triangulation positioning accuracy of \(0.44\;\mathrm{m}\) can be achieved using the DOAs estimated by the proposed method from only two TRPs, which meets the 3GPP R17 requirements for commercial use cases \cite{3gpp.38.857}.

\subsection{Paper Outline and Notations}
This paper is organized as follows. First, the system model for 5G-signal-based positioning is presented and the problem of JADE is formulated in Section \ref{sec:sys-model}. It is followed by real-data-based array modeling error analysis in Section \ref{sec:array-modeling-error}. Then the JADE signal model in the presence of array modeling errors is re-formulated in this section. Afterwards, details of the proposed array modeling error calibration and JADE scheme are presented in Section \ref{sec:proposed-method}. Its computational complexity and storage requirement are also analyzed. Next, in Section \ref{sec:impl-issu}, enhancements to the proposed scheme are presented for further complexity reduction, which yields an efficient implementation for the proposed method. In Section \ref{sec:perf-eval}, in-depth performance evaluations, including numerical simulation and field tests, are provided. Furthermore, Section \ref{sec:ext} presents discussions regarding the adaptability of the proposed method to other potential scenarios. Finally, Section \ref{sec:conclusion} concludes the paper.

In the rest of this paper, vectors and matrices are denoted by boldface lowercase and boldface uppercase letters, respectively, where vectors are by default in column orientation. Italic English letters and lowercase Greek letters denote scalars. Blackboard-bold characters denote number sets, in particular, \(\mathbb{R}\) and \(\mathbb{C}\) represent the sets of real and complex numbers, respectively. \(\jmath = \sqrt{-1}\) denotes the imaginary unit. \(x = \mathcal{O}(a)\) for \(a > 0\) denotes that \(\exists k_{1}, k_{2}>0\), such that \(k_{2} \cdot a \leq x \leq k_{1} \cdot a\). TABLE \ref{tab:notations} lists all the other notations used in this paper and their meanings.

\begin{table}[htbp]
  \caption{Notations}
  \begin{center}
    \begin{tabular}{cl}
      \hline\hline\tstrut
      \((\cdot)^{\mathrm{T}}\) & transpose of a vector or matrix \\
      \((\cdot)^{\mathrm{H}}\) & conjugate transpose of a vector or matrix \\
      \((\cdot)^{-1}\) & inverse of a square matrix \\
      \(|a|\) & absolute value of the scalar \(a\) \\
      \(|\mathbf{a}|\) & a vector with absolute values of entries of vector \(\mathbf{a}\) \\
      \(\left[\mathbf{a}\right]_n\) & \(n\)-th element of vector \(\mathbf{a}\) \\
      \(\left[\mathbf{A}\right]_{m,n}\) & element at \(m\)-th row and \(n\)-th column of matrix \(\mathbf{A}\) \\
      \(\left[\mathbf{a}\right]_{i:j}\) & subvector formed by \(i\)-th to \(j\)-th entries \\
      \(\mathbf{0}_N\) & a zero vector of length \(N\) \\
      \(\odot\) & the Hadamard (element-wise) matrix product \\
      \(\mathcal{U}\left(a, b\right]\) & uniform distribution from \(a\) to \(b\) \\
      \(\mathrm{diag}(\mathbf{a})\) & diagonal matrix whose main diagonal entries are \(\mathbf{a}\) \\
      \(\mathsf{FFT}[\mathbf{a}, P]\) & \(P\)-point FFT on \(\mathbf{a}\) \\
      \(\mathsf{Toeplitz}(\mathbf{a})\) & a Hermitian Toeplitz matrix with elements of \(\mathbf{a}\) \\
      \(a(x)\) & a scalar-valued function with input variable of \(x\) \\
      \(\mathbf{a}(x)\) & a vector-valued function with input variable of \(x\) \\
      \hline\hline\bstrut
    \end{tabular}
    \label{tab:notations}
  \end{center}
\end{table}

\section{System Model}
\label{sec:sys-model}
\begin{figure}[htb]
  \centering
  \includegraphics[width=0.48\textwidth]{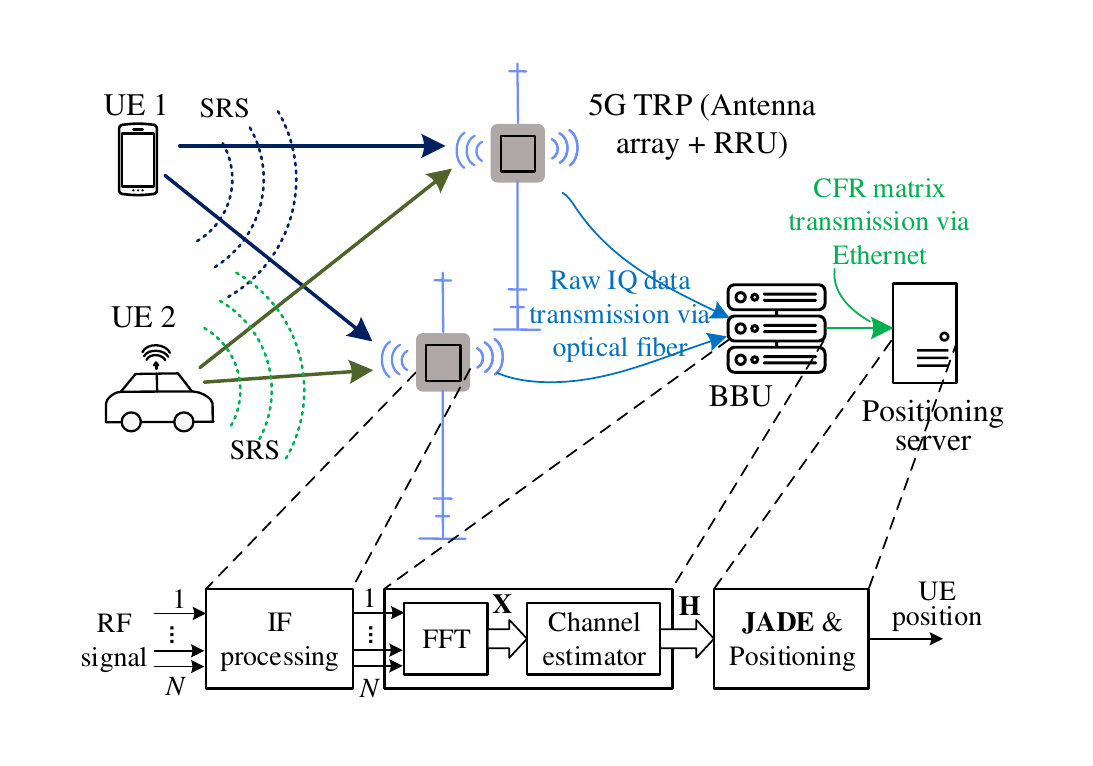}
  \caption{Positioning scenario based on 5G cellular network with a picocell gNB. The abbreviations of BBU, RRU, IF and IQ data stand for baseband unit, radio remote unit, intermediate frequency and in-phase and quadrature data, respectively.}
  \label{fig:localization_scenario}
\end{figure}

As shown in Fig. \ref{fig:localization_scenario}, we consider the positioning scenario in which the 5G TRPs receive the up-link SRS transmitted by the user equipment (UE) and determine the UE position by jointly estimating the DOA and TOA. The SRS is an OFDM signal with a pre-defined Zadoff-Chu sequence for each unique terminal \cite{3gpp.38.211}. It is designed for uplink channel sounding, while the sounding result in the frequency domain, which is also denoted as the channel frequency response (CFR) of the wireless channel, contains the DOA and TOA information of both the direct path and reflected paths.

In this section, we will setup the multichannel CFR-based JADE signal model based on the following hypothesis:
\begin{enumerate}
  \item Since the main focus of this paper is 2-D parameter estimation for positioning in a 2-D space, we assume that the UE antenna and the TRP antennas are at the same height and the TRP antenna arrays are all horizontally placed. Therefore, the DOA refers to the azimuth angle and is defined as the angle with respect to the array broadside in this paper. 
\item We only consider the problem of positioning a single user terminal, while the positioning of multiple terminals can be easily decomposed into this case owing to the orthogonality between uplink SRS sequences from different terminals.
\item We consider the problem of JADE using a single SRS symbol, i.e. JADE in a single measurement vector (SMV) scenario, which is more promising for moving UE positioning. Therefore, the time index is omitted in the following signal model.
\end{enumerate}

Considering an SRS symbol which occupies \(M\) subcarriers
is transmitted by the UE and impinges on a 5G TRP which is equipped with an array of \(N\) antenna elements. The received time-domain multichannel SRSs are first transformed to the frequency-domain via the FFT. The frequency-domain received signal of channel \(n\) is denoted as \(\mathbf{x}_n = \left[X_{1,n}, X_{2,n},\dots,X_{M,n}\right]^{\mathrm{T}}\in\mathbb{C}^{M\times1}\), in which \(X_{m,n}\) represents the SRS component received at \(n\)-th receiving channel and \(m\)-th subcarrier. The frequency-domain signals from all channels are arranged into a single matrix \(\mathbf{X} = \left[\mathbf{x}_1,\mathbf{x}_2,\dots,\mathbf{x}_{N}\right]\in\mathbb{C}^{M\times N}\).

Assuming that the SRS arrives at the TRP via \(K\) paths with DOAs of \(\{\tilde{\theta}_k\}_{k=1}^K\), TOAs of \(\{\tilde{\tau}_k\}_{k=1}^K\), and attenuation coefficients of \(\{\tilde{\gamma}_k\}_{k=1}^K\),
and denoting the index of the LOS path as \(k_{\mathrm{LOS}}\), then the received signal matrix can be represented as
\begin{equation}
\mathbf{X} = \sum\limits_{k = 1}^K \left[{{{\tilde \gamma }_k}{\mathbf{S}} \cdot {{\mathbf{a}}_\tau }({{\tilde \tau }_k}){\mathbf{a}}_\theta^{\mathrm{T}}({{\tilde \theta }_k})}\right] + {\mathbf{W}}. \label{eq:matrixX}
\end{equation}

Symbols used in equation (\ref{eq:matrixX}) are clarified as follows. First, \(\mathbf{S} = \mathrm{diag}\left(\left[S[1], S[2], \dots, S[M]\right]^{\mathrm{T}}\right)\) is the SRS data matrix, where \(S[m], m = 1,2,\dots, M\) is the transmitted SRS sequence.

Next, \(\mathbf{a}_{\tau}(\cdot): \mathbb{T}\to\mathbb{C}^{M\times1}\) is the delay signature function which represents the frequency-domain structure of the SRS. It is a known function whose input is the path delay, or the TOA of the signal, and the output is the delay signature vector. The interested range of delay \(\mathbb{T}\subseteq\mathbb{R}\) can be determined by the power coverage of the TRP. For example, assuming that \(\tau_\text{min}\) and \(\tau_{\text{max}}\) represent the minimum and maximum measurable path delay, respectively, then \(\mathbb{T} = \left\{\tau\big| \tau_{\text{min}}\le\tau\le\tau_{\text{max}}\right\}\). The \(m\)-th element of the output delay signature vector \(\mathbf{a}_{\tau}(\tilde{\tau}_k)\) represents the shift factor caused by the path delay \(\tilde{\tau}_k\) in the \(m\)-th subcarrier, and can be denoted as
\begin{equation}
  \left[\mathbf{a}_{\tau}(\tilde{\tau}_k)\right]_m = \exp\left(-\jmath2\pi(m-1)\Delta f\tilde{\tau}_k\right),\quad m = 1,2,\dots, M,
  \label{eq:a_tau}
\end{equation}
where \(\Delta f\) is the subcarrier spacing.

Then, \(\mathbf{a}_{\theta}(\cdot): \mathbb{U}\to\mathbb{C}^{N\times1}\) is the steering-vector function determined by the spatial structure of the antenna array. It is also a known function with DOA as its input and steering-vector at that specific direction as the output. The interested range of DOA \(\mathbb{U}\subseteq\mathbb{R}\) is determined by the angular coverage of the TRP antenna array. Also as an example, when \(\theta_{\text{min}}\) and \(\theta_{\text{max}}\) are the minimum and maximum measurable angles, respectively, we have \(\mathbb{U} = \left\{\theta\big| \theta_{\text{min}}\le\theta\le\theta_{\text{max}}\right\}\). Assuming that the receiving array is an ideal uniform linear array (ULA), then the \(n\)-th element of the output steering-vector for the impinged DOA of \(\tilde{\theta}_k\) is
\begin{equation}
  \left[\mathbf{a}_{\theta}\left(\tilde{\theta}_k\right)\right]_n=\exp\left(\jmath2\pi\frac{(n-1)d\sin\tilde{\theta}_k}{\lambda}\right),\quad n = 1,2,\dots,N,
  \label{eq:eq:a_theta}
\end{equation}
where \(d\) is the array element spacing and \(\lambda\) is the wavelength.

Lastly, matrix \(\mathbf{W}\in\mathbb{C}^{M\times N}\) in equation \eqref{eq:matrixX} is the noise matrix, whose element \(\left[\mathbf{W}\right]_{m,n}\) represents the noise component at \(n\)-th receiving channel and \(m\)-th subcarrier.

As shown in Fig. \ref{fig:localization_scenario}, after the FFT operation, a channel estimator is followed in the BBU to estimate the wireless channel response. Assuming that the least-squares (LS) algorithm is applied using the a prior of the SRS sequence, the estimated CFR matrix \(\mathbf{H}\in\mathbb{C}^{M\times N}\) can be derived as
\begin{equation}
  \mathbf{H} = \mathbf{S}^{-1}\mathbf{X} = \sum\limits_{k = 1}^K \left[{{{\tilde \gamma }_k} \cdot {{\mathbf{a}}_\tau }({{\tilde \tau }_k}){\mathbf{a}}_\theta^{\mathrm{T}}({{\tilde \theta }_k})}\right] + \mathbf{W}',
  \label{eq:Hmatrix}
\end{equation}
where \(\mathbf{W}' = \mathbf{S}^{-1}\mathbf{W}\in\mathbb{C}^{M\times N}\) represents the noise components in the CFR matrix \(\mathbf{H}\).

The basic signal model for the JADE problem addressed in this paper is expressed by equation \eqref{eq:Hmatrix} and is formulated as:
\begin{quote}
  Given a single space-frequency CFR matrix \(\mathbf{H}\), estimate the LOS DOA \(\tilde{\theta}_{k_{\mathrm{LOS}}}\) and TOA \(\tilde{\tau}_{k_\mathrm{LOS}}\).  
\end{quote}

 This ideal signal model without considering the impairments induced by the array modeling errors is quite similar to the JADE models for Wi-Fi \cite{kotaru2015_SpotFiDecimeterLeva}, UWB \cite{hua2016_JointEstimationDOA}, GSM \cite{vanderveen1997_JointAngleDelay}, or LTE \cite{shamaei2021_JointTOADOAa} signals in the literature. The impact of array modeling errors of picocell gNBs will be analyzed and will be incorporated into the signal model in Section \ref{sec:array-modeling-error}.

\section{Array Modeling Error Analysis Based on Real-Data}
\label{sec:array-modeling-error}

\begin{figure}[htb]
  \centering
  \includegraphics[width=0.26\textwidth]{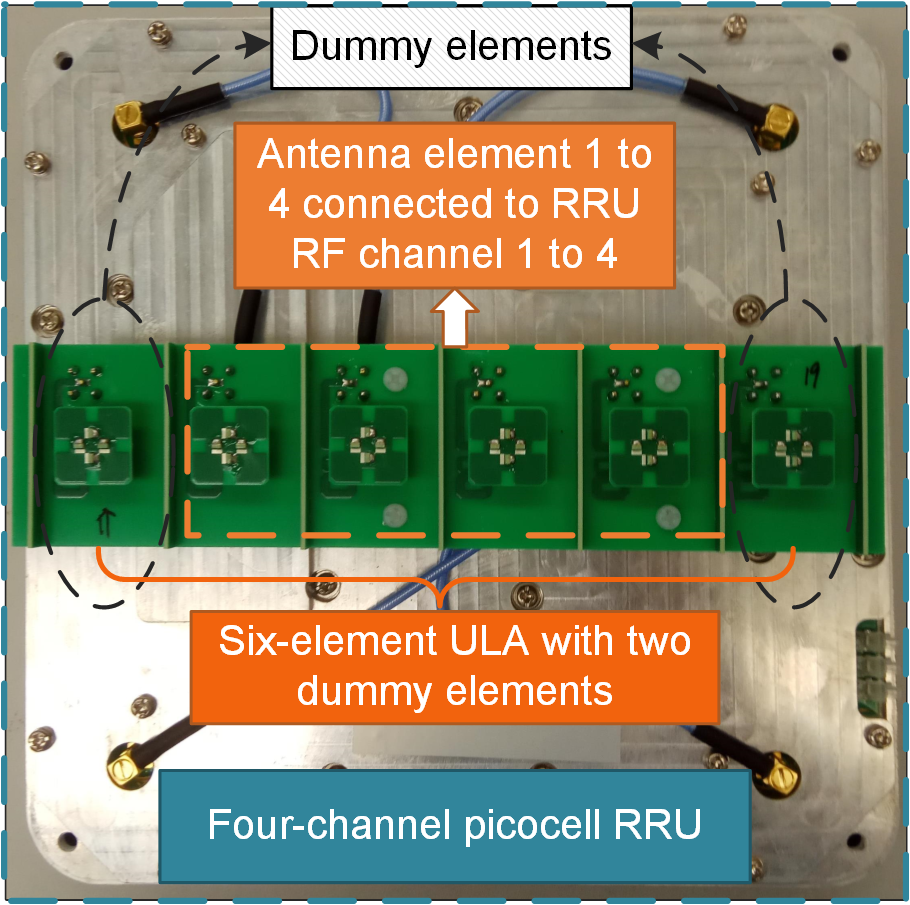}
  \caption{RRU and antenna array of the 5G picocell gNB employed in experiments of this paper.}
  \label{fig:rru_antenna}
\end{figure}

The array modeling errors can be roughly divided into two parts: the part induced by the multichannel receiver and the part induced by the antenna elements, which are referred as RF channel errors and antenna errors in this paper, respectively. The RRU and the antenna array of a picocell gNB employed in
experiments of this paper
are evaluated for exemplifying the spatial-frequential characteristics of these errors. The antenna array is a six-element ULA working at 5G NR frequency range 1 (FR1) with an inter-element spacing of \(3\;\mathrm{cm}\), as shown in Fig. \ref{fig:rru_antenna}. Among these six antenna elements, those two at both sides are dummy elements which are designed for alleviating the mutual coupling effect of the whole array and are not connected to any RF channels; while the middle four antennas are valid elements for conducting the receiving signals to the according RF channels.

\subsection{RF Channel Error Analysis}
\label{sec:rf-channel-error}
Wideband SRS is conducted to the receiving RF channels of this four-channel picocell RRU directly via coaxial cables and a RF power divider to measure the RF channel errors, which is also referred as the amplitude and phase responses of the RF channels.
The SRS is configured according to TABLE \ref{tab:srs_parameters},
which is also the default SRS configuration used throughout this paper.

\begin{table}[htbp]
  \caption{Waveform parameters for uplink SRS assumed in this study}
  \begin{center}
    \begin{tabular}{cc}
      \toprule
      \bf{Parameter} & \bf{Value} \\
      \midrule
      Working frequency range & FR1 (\(4.80 - 4.90\;\mathrm{GHz}\)) \\
      Numerology \(\mu\) & \(1\) \\
      Subcarrier spacing & \(30\;\text{kHz}\) \\
      Number of resource blocks & \(272\) \\
      Number of subcarriers & \(3264\) \\
      SRS pattern & Comb-two \cite{3gpp.38.211} \\
      SRS transmission bandwidth & \(100\;\text{MHz}\) \\
      SRS periodicity & \(80\;\text{ms}\) \\
      Cyclic prefix mode & Normal \\
      FFT size & \(4096\) \\
      Sampling rate & \(122.88\;\mathrm{Msps}^{\mathrm{a}}\) \\
      \bottomrule
      \multicolumn{2}{l}{\(^{\mathrm{a}}\mathrm{sps}\): abbreviation for symbols per second.}\\
    \end{tabular}
    \label{tab:srs_parameters}    
  \end{center}
\end{table}

\begin{figure}[htb]
  \centering
  \includegraphics{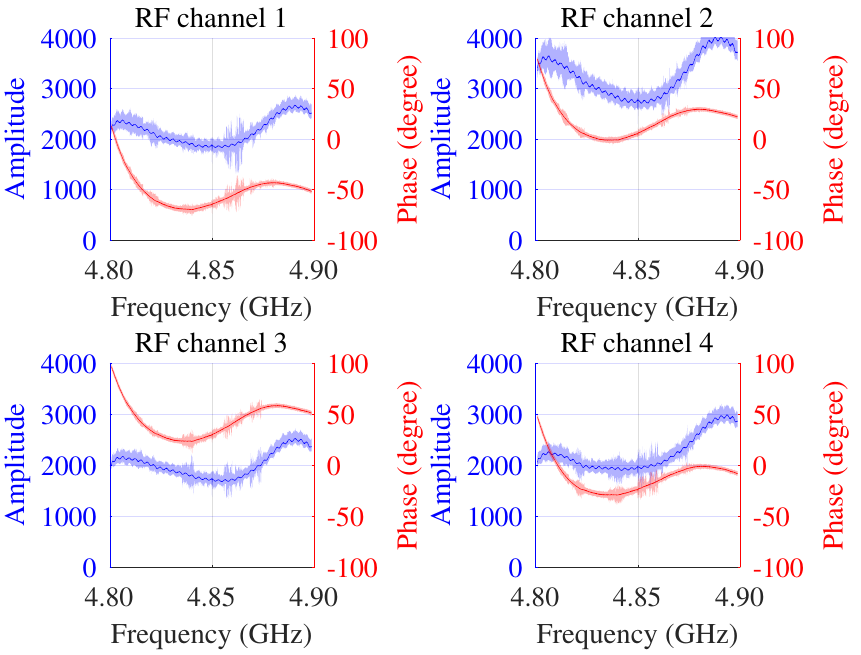}
  \caption{RF channel error measurements of the four-channel RRU.}
  \label{fig:channel_errors}
\end{figure}

The measured channel errors are shown in Fig. \ref{fig:channel_errors}, where in each subfigure, the solid lines are the average values of \(500\) repeated RF channel amplitude and phase measurements and the shadow areas illustrate the upper and lower bounds of those measurements. It can be inferred from Fig. \ref{fig:channel_errors} that: First, the initial phase of the channel responses differ distinctly, which will lead to the DOA estimation bias; Second, in the frequency-domain, the non-linear components predominate, which will distort the signal spectrum in the TOA domain.

\subsection{Antenna Error Analysis}
\label{sec:ant-error}
Different from RF channel errors, owing to the mutual coupling effects, location perturbations and beampattern errors of antenna elements, the antenna errors are direction-dependent. As the DOA is obtained from phase shifts across receiving antennas, this paper mainly focuses on the antenna phase errors.

\begin{figure}[htb]
  \centering
  \includegraphics[width=0.45\textwidth]{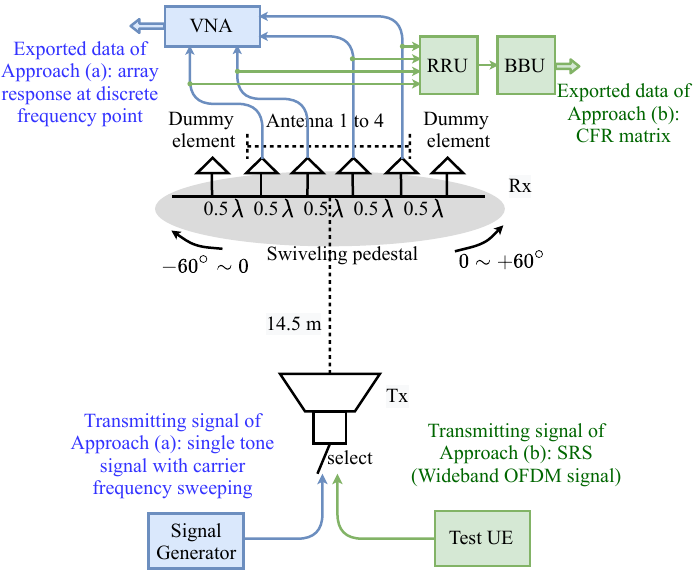}
  \caption{Setups for anechoic chamber experiments, where blue and green colors mark the components used for array response measuring (Section \ref{sec:ant-error}) and for field test in a multipath-free environment (Section \ref{sec:anechoic-exp}), respectively.}
  \label{fig:chamber_setups}
\end{figure}

Phase responses of the four-element antenna array of the picocell gNB are measured in a far-field anechoic chamber by Approach (a) shown in Fig. \ref{fig:chamber_setups} (Approach (b) is used to collect CFR matrices for performance evaluation, as will be presented later in Section \ref{sec:anechoic-exp}).

Here, single tone signals are transmitted by the horn antenna with center frequency sweeping from \(4.80\;\mathrm{GHz}\) to \(4.90\;\mathrm{GHz}\) with a step size of \(10\;\mathrm{MHz}\). The receiving arrays are placed on a swiveling pedestal, which swivels from \(-60^\circ\) to \(+60^\circ\) in an angular step of \(5^\circ\). Phase responses of this antenna array are then measured by a vector network analyzer (VNA), from which the phase shifts caused by free-space propagations are subtracted,
deriving the desired phase errors caused by the imperfect array response.

\begin{figure}[htb]
  \centering
  \subfloat[Antenna phase errors across frequencies in different incident angles.]{\includegraphics[width=0.25\textwidth]{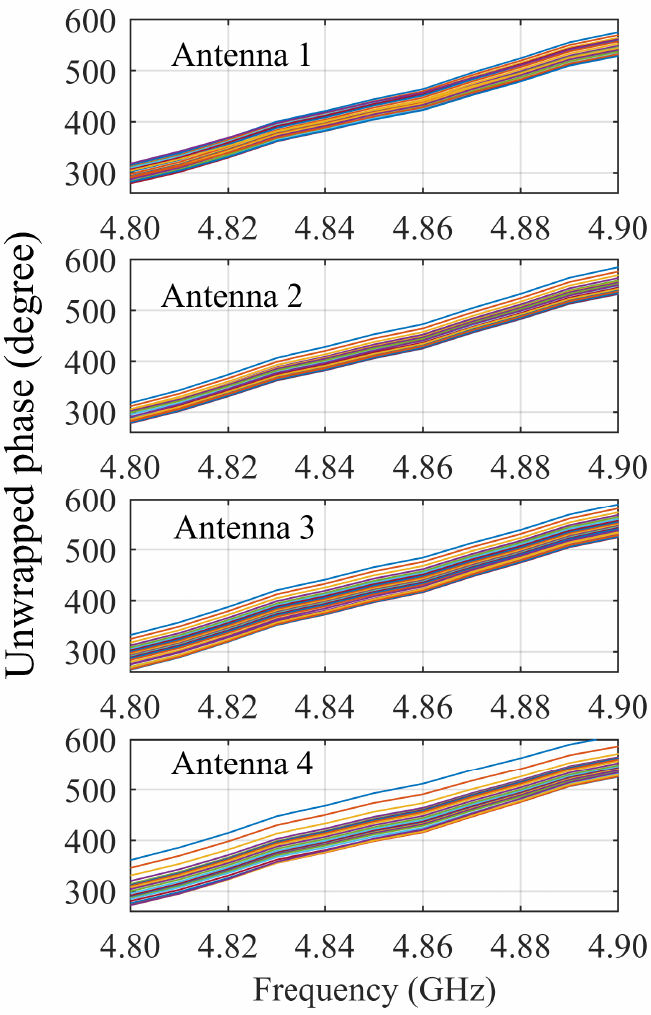}}
  \hfill
  \subfloat[Histogram of distance bias caused by antenna errors.]{\includegraphics[width=0.231\textwidth]{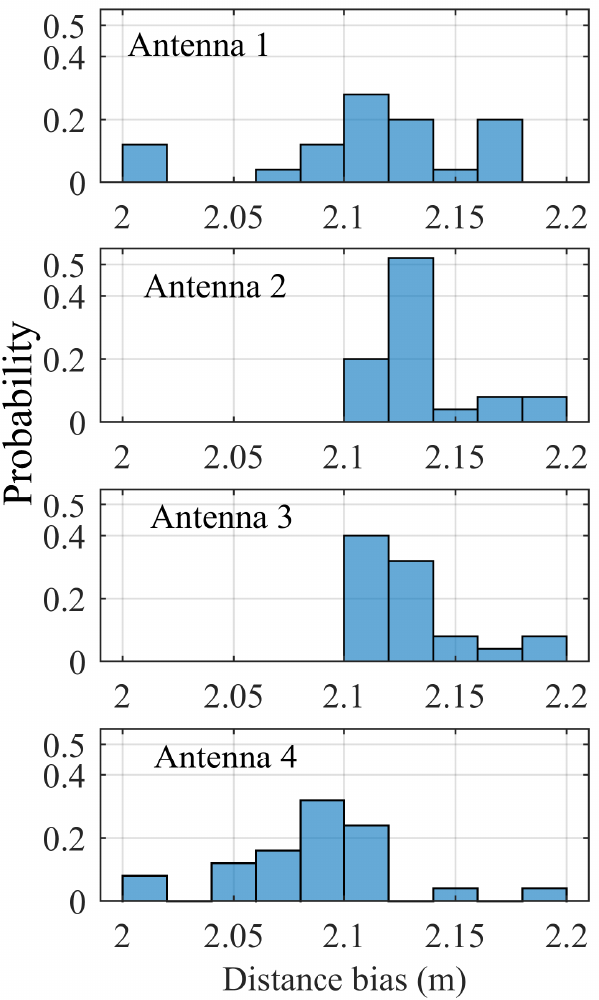}}
  \caption{Antenna phase error characteristics in the frequency-domain.}
  \label{fig:phase_frequency}
\end{figure}

Antenna phase error is first investigated in the frequency domain in Fig. \ref{fig:phase_frequency}. Different lines in each subfigure of Fig. \ref{fig:phase_frequency}(a) illustrate the phase errors of a corresponding antenna element when the signal impinges on the array from different directions. It demonstrates that, oppositing to the phase errors caused by RF channels, those caused by antenna elements are dominated by linear components. This phenomenon is plausible since the antenna element proper is a passive device which merely conducts the received wireless signal to the RF port. Moreover, these phase error curves shown in Fig. \ref{fig:phase_frequency}(a) are nearly parallel to each other. By applying a linear transformation to the slopes of these curves according to the following equation:
\begin{equation}
\text{Distance bias} = \frac{\text{Slope [rad/Hz]}\cdot{}c}{2\pi},
\end{equation}
in which \(c\) presents the speed of light in vacuum, the distributions of the distance biases caused by the antenna errors when the incident direction varies are revealed in Fig. \ref{fig:phase_frequency}(b).
It can be inferred that the distance biases are all in the range of \([2.0, 2.2]\text{m}\), with the variance across antennas and across incident directions less than \(0.2\text{m}\), which is much less than the distance resolution of \(3\text{m}\). This means that, the antenna errors cause a nearly identical TOA shift in all receiving channels for signals impinging from any direction.


\begin{figure}[htb]
  \centering
  \includegraphics[width=0.35\textwidth]{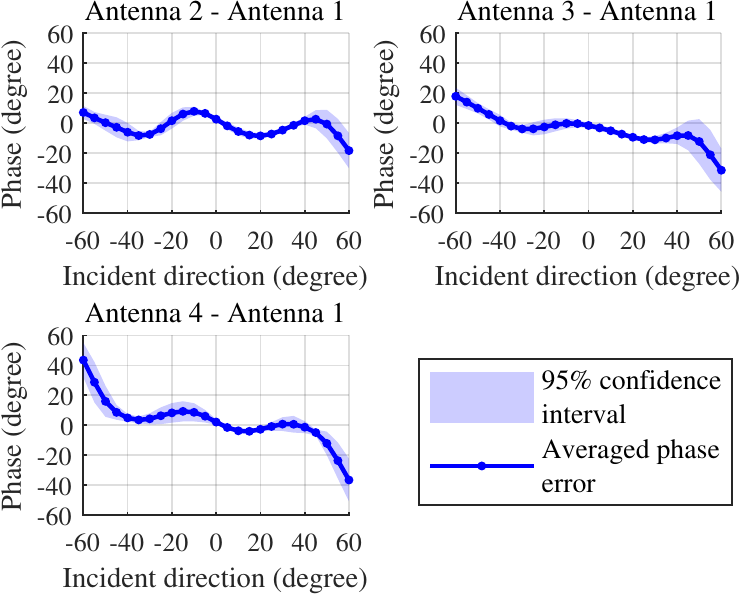}
  \caption{Antenna phase errors relative to the first antenna element.}
  \label{fig:phase_differences_angles}
\end{figure}

Then the antenna phase errors are investigated in the angular domain in Fig. \ref{fig:phase_differences_angles}. Offsets of these errors relative to the first antenna element averaged by measurements sampled at different frequencies are shown in Fig. \ref{fig:phase_differences_angles}, with \(95\%\) confidence intervals of these measurements also highlighted by the shadowed areas.
It demonstrates that, to the contrary of channel errors, variances of antenna phase errors across frequencies are relatively small and the antenna phase error is highly direction-dependent. The consequences are two-fold: First, to reduce the complexity of calibration, the antenna phase error can be regarded as a constant across different frequencies and only spatial domain calibration is necessary; Second, since the incident direction of the SRS is the parameter to be estimated for uplink positioning, which is unknown to the gNB, the gNB is unable to determine the calibration coefficients for antenna errors.

\subsection{Signal Model Refinement in the Presence of Array Modeling Errors}
\label{sig-model-refinement}
Based on the aforementioned analysis, the signal model of equation (\ref{eq:Hmatrix}) is revised by incorporating the array modeling errors, including both the RF channel errors and the antenna errors. The frequency-selective RF channel errors are modeled as a matrix \(\mathbf{\Gamma}\in\mathbb{C}^{M\times N}\), whose element \(\left[\boldsymbol{\Gamma}\right]_{m,n}\) represents the amplitude and phase response of the RF channel in the \(m\)-th subcarrier and in the \(n\)-th receiving channel.
Then the direction-dependent antenna error is formulated as a vector-valued function \(\boldsymbol{\zeta}(\theta): \mathbb{U}\to\mathbb{C}^{N\times 1}\), which outputs the antenna errors
as a \(N\)-dimensional vector
when a DOA value is given. The \(n\)-th element of the output vector is \(\left[\boldsymbol{\zeta}(\theta)\right]_n = \exp(\jmath\phi_n(\theta))\), where \(\phi_n(\theta): \mathbb{U} \to [-\pi,\pi]\) is the phase error function of the \(n\)-th antenna element. 
By this means, all kinds of errors
are gathered into a single direction-dependent antenna error function and thereby equation (\ref{eq:Hmatrix}) can be extended to
\begin{equation}
  \mathbf{H} = \sum\limits_{k = 1}^K \left[{{{\tilde \gamma }_k} \cdot {{\mathbf{a}}_\tau }({{\tilde \tau }_k}){\mathbf{a}}_\theta'^{\mathrm{T}}({{\tilde \theta }_k})}\right]\odot\boldsymbol{\Gamma} + \mathbf{S}^{-1}\mathbf{W}',
  \label{eq:revised_H}
\end{equation}
where \(\mathbf{a}_\theta'(\cdot)\) represents the actual spatial steering-vector function. According to the aforementioned antenna error model, it relates to the ideal steering-vector function and the antenna error function as follows:
\begin{equation}
  \label{eq:a_impaired}
  \mathbf{a}_\theta'(\theta) = \mathbf{a}_\theta(\theta)\odot\boldsymbol{\zeta}(\theta).
\end{equation}

Finally, the problem addressed in this paper is concluded as:
\begin{quote}
  Given a single estimated space-frequency CFR matrix \(\mathbf{H}\) and the pre-measured data set for the array modeling error, estimate the desired positioning parameters \(\tilde{\theta}_{k_{\mathrm{LOS}}}\) and \(\tilde{\tau}_{k_{\mathrm{LOS}}}\). 
\end{quote}

\section{Array modeling error calibration and JADE}
\label{sec:proposed-method}

\begin{figure*}[htb]
  \centering
  \includegraphics{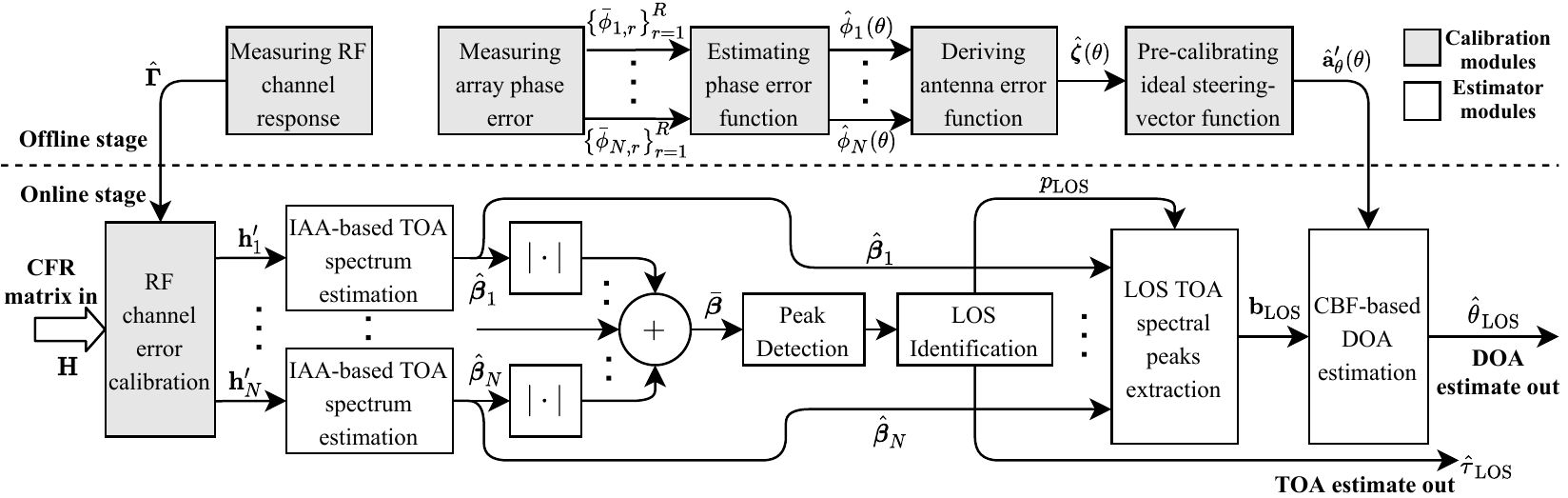}
  \caption{Flowchart of the proposed JADE scheme.}
  \label{fig:proposed_method_flowchart}
\end{figure*}

The flowchart of the proposed JADE scheme
is shown in Fig. \ref{fig:proposed_method_flowchart}.
The proposed scheme is composed of two main parts: calibration modules which cope with the RF channel errors and the antenna errors separately and estimator modules consisting of an IAA-based TOA spectral analyzer and a CBF.


\subsection{Calibration Modules}
As shown in Fig. \ref{fig:proposed_method_flowchart}, each entry of the received multichannel CFR matrix is first divided by the measured RF channel response matrix \(\hat{\mathbf{\Gamma}}\) to calibrate the RF channel phase and amplitude inconsistency. The element at \(m\)-th row and \(n\)-th column of the CFR matrix after calibrating the channel errors is 
\begin{equation}
  \left[\mathbf{H}'\right]_{m,n} = \left[\mathbf{H}\right]_{m,n} / \left[\hat{\mathbf{\Gamma}}\right]_{m,n}.
  \label{eq:channel_calibration}
\end{equation}

On the contrary, the direction-dependent antenna error prevents the receiver from calibrating it directly in the received CFR.
Therefore, in the proposed scheme, the steering-vector function is calibrated at offline stage to counteract the antenna error.
Specifically, there are four main steps in the offline stage for antenna error calibration as illustrated by the upper part of Fig. \ref{fig:proposed_method_flowchart}.
First, antenna phase errors in \(R\) discrete angles within the array coverage are measured using the approach illustrated in Section \ref{sec:ant-error}. It is noteworthy to mention that, electromagnetic numerical simulations can be considered to reduce the overheads of measuring.
The angular sampling set and the phase error measurements for antenna element \(n\) on this set are denoted as \(\left\{\bar{\theta}_r\right\}_{r=1}^R\) and \(\left\{\bar{\phi}_{n,r}\right\}_{r=1}^R, n = 1,\dots,N\), respectively.

Then the phase error function of each antenna element is estimated based on measured data sets. This function estimation problem can be formulated as a regression problem, for which LS-based polynomial curve fitting, support vector machine, neural network, or other regression method, can be applied here.
Take the naive polynomial curve fitting method as an example. The estimated function for the \(n\)-th antenna element has the form of
\begin{equation}
  \phi_{\text{poly}, n}(\theta, \mathbf{g}) = \sum_{i=1}^Ig_{i}\theta^{i-1}, \quad n = 1, \dots, N,
\end{equation}
where \(\mathbf{g}\in\mathbb{R}^{I\times 1}\) presents the coefficient vector of a \(I\)-th-order polynomial. According to the LS curve fitting principle, the phase error function can be determined by
\begin{equation}
  \begin{cases}
    \hat{\phi}_n(\theta) = \phi_{\text{poly}, n}(\theta, \hat{\mathbf{g}}), \\
    \hat{\mathbf{g}} = \arg\min_{\mathbf{g}}\sum_{r=1}^R\left|\phi_{\text{poly},n}(\bar{\theta}_r, \mathbf{g}) - \bar{\phi}_{n,r}\right|^2.
  \end{cases}  
\end{equation}

\begin{figure}[htb]
  \centering
  \subfloat{\includegraphics[width=0.38\textwidth]{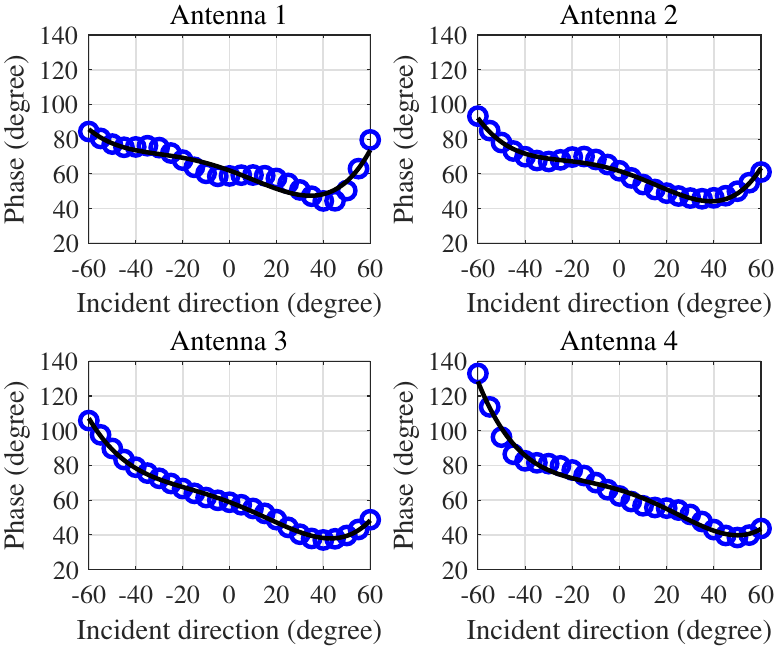}} \\
  \subfloat{\includegraphics[width=0.31\textwidth]{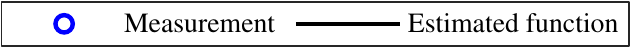}}
  \caption{Polynomial curve fitting for antenna phase error function estimation based on real-measured phase error data set.}
  \label{fig:antenna_function_estimation}
\end{figure}

For the antenna array shown in Fig. \ref{fig:rru_antenna}, whose phase error measurements have been demonstrated in Fig. \ref{fig:phase_frequency} and Fig. \ref{fig:phase_differences_angles}, its phase error functions for all elements estimated by fourth-order polynomial curve fitting are illustrated in Fig. \ref{fig:antenna_function_estimation}.

After that, based on the estimated phase error function \(\hat{\phi}_n(\theta), n = 1,\dots, N\), the antenna error function \(\boldsymbol{\zeta}(\theta)\) is determined, with the \(n\)-th element of the function output for signal impinging from the direction of \(\theta\) equals to
\begin{equation}
  \left[\hat{\boldsymbol{\zeta}}(\theta)\right]_n = \exp(\jmath\hat{\phi}_n(\theta)).
\end{equation}

Lastly, the ideal steering-vector function is pre-calibrated according to the estimated antenna error function as follows:
\begin{equation}
  \hat{\mathbf{a}}_{\theta}'(\theta) = \mathbf{a}_{\theta}(\theta)\odot\hat{\boldsymbol{\zeta}}(\theta).
\end{equation}

This offline-stage pre-calibrated steering-vector function \(\hat{\mathbf{a}}_{\theta}'(\theta)\) captures the antenna errors and delineate the actual responses of the antenna array for signals impinging from different directions more precisely than the ideal steering-vector function. It is used by the estimator modules at online-stage to improve the DOA estimation accuracy.  

\subsection{Estimator Modules}
\label{sec:proposed_jade}

Conventional joint DOA-TOA estimators usually incorporate 2-D searching on the DOA-TOA plane or operations on the huge-dimensional space-frequency data matrix, which are both computational intensive. In the proposed scheme shown in Fig. \ref{fig:proposed_method_flowchart}, TOA and DOA are estimated cascadingly with an IAA-based spectral estimator and a CBF. The main idea behind such a design is to fully exploit the abundant subcarrier resources of the 5G signal and the super-resolution ability of the IAA spectral estimator to separate multipaths by their delays, then a naive CBF can be applied to the segregated multipath components to retrieve their DOAs.

The reason of employing the IAA method for TOA spectrum estimation is three-fold:
\begin{enumerate}
\item Compared with parametric spectral estimators which need to tune their parameters for desired performance \cite{kotaru2015_SpotFiDecimeterLeva, shamaei2021_JointTOADOAa, gong2017_RobustIndoorWirelea, fan2021_LocalizationBasedI}, the IAA method is a nonparametric estimator which can be employed for data set from any scenario without cumbersome parameter-tuning;
\item Compared with extensively used subspace-based spectral estimators, which demand multiple snapshot or need to be combined with the smoothing technique to construct multiple measurements artificially \cite{kotaru2015_SpotFiDecimeterLeva, bazzi2016_SpatiofrequentialSm}, the IAA method can be applied in a SMV scenario directly;
\item Compared with the pseudo-spectrum obtained by the MUSIC algorithm, whose spectral values do not represent signal amplitude, the IAA method is able to recover the amplitude and phase of each signal component, which benefits the subsequent LOS discriminator and can be directly used by the following DOA estimator.
\end{enumerate}


According to equation \eqref{eq:revised_H} and \eqref{eq:channel_calibration}, the \(n\)-th column of the channel-error calibrated CFR matrix \(\mathbf{H}'\) can be derived as
\begin{equation}
  \mathbf{h}_n' = \sum_{k=1}^{K}\tilde{\beta}_{n,k}\mathbf{a}_\tau^{\mathrm{T}}(\tilde{\tau}_k) + \mathbf{w}_n,
  \label{eq:hn1}
\end{equation}
where \(\mathbf{w}_n\in\mathbb{C}^{M\times 1}\) represents the noise components in \(\mathbf{h}_n'\) and \(\tilde{\beta}_{n,k} = \tilde{\gamma}_k\cdot\alpha_n(\tilde{\theta}_k)\), in which \(\alpha_n(\tilde{\theta}_k) = \left[\mathbf{a}_\theta'(\tilde{\theta}_k)\right]_n\).

IAA method is a dense spectrum estimator which estimates the power spectrum at a predefined grid set, hence the signal model expressed by equation \eqref{eq:hn1} is further extended to make the following explanation concise. Suppose the TOA span of \(\mathbb{T}\) is covered by a uniform grid set \(\{\tau_p\}_{p=1}^P\), and the attenuation coefficient sounded at the \(p\)-th TOA grid by the \(n\)-th channel is denoted as \(\beta_{n,p}, p = 1, \dots, P, n = 1, \dots, N\). Then \(\beta_{n,p}\) has the form of:
\begin{equation}
  \beta_{n,p} = \begin{cases}
    \tilde{\beta}_{n,k}, \quad\tau_p = \tilde{\tau}_k, \\
    0,\quad\text{elsewhere}.
  \end{cases}
\end{equation}

Then equation \eqref{eq:hn1} can be reformulated as
\begin{equation}
  \mathbf{h}_n' = \mathbf{A}_{\tau}\boldsymbol{\beta}_n + \mathbf{w}_n,
  \label{eq:hn2}
\end{equation}
where \(\boldsymbol{\beta}_n = \left[\beta_{n,1},\beta_{n,2},\dots,\beta_{n,P}\right]^{\mathrm{T}}\in\mathbb{C}^{P\times 1}\) is the attenuation coefficient vector of the \(n\)-th channel and \(\mathbf{A}_{\tau} = \left[\mathbf{a}_\tau(\tau_1),\dots,\mathbf{a}_\tau(\tau_P)\right]\in\mathbb{C}^{M\times P}\) is the over-complete delay signature matrix.

In the IAA framework, solution for each element of \(\boldsymbol{\beta}_n\) is sought according to the weighted least squares criteria \cite{yardibi2010_SourceLocalization}:
\begin{equation}
  \hat{\beta}_{n,p} = \arg\min_{\beta_{n,p}}\left\|\mathbf{h}_n' - \beta_{n,p}\mathbf{a}_{\tau}(\tau_p)\right\|_{\mathbf{R}_{n,p}'^{-1}}^2,\quad p = 1,\dots,P,
  \label{eq:object_fun}
\end{equation}
in which \(\left\|\mathbf{x}\right\|_{\mathbf{Q}}^2=\mathbf{x}^{\mathrm{H}}\mathbf{Q}\mathbf{x}\) represents the square of the weighted \(\ell_2\)-norm of the vector \(\mathbf{x}\). In \eqref{eq:object_fun}, the weighted matrix \(\mathbf{R}_{n,p}'^{-1}\) is the inverse of the interference covariance matrix \(\mathbf{R}_{n,p}'\), while the interference here refers the signal components arrived the receiver at delays other than the delay of current interest \(\tau_p\). Therefore, \(\mathbf{R}_{n,p}'\) can be presented by
\begin{equation}
  \mathbf{R}_{n,p}' = \mathbf{R}_n -\left|\beta_{n,p}\right|^2\mathbf{a}_{\tau}(\tau_p)\mathbf{a}_{\tau}^{\mathrm{H}}(\tau_p),
  \label{eq:R_interfere}
\end{equation}
where \(\mathbf{R}_n\) represents the covariance matrix of the signal received by the \(n\)-th channel.

Solution to equation \eqref{eq:object_fun} is derived to be 
\begin{equation}
  \hat{\beta}_{n,p} = \frac{\mathbf{a}_{\tau}^{\mathrm{H}}(\tau_p)\mathbf{R}_{n,p}'^{-1}\mathbf{h}_n'}{\mathbf{a}_{\tau}^{\mathrm{H}}(\tau_p)\mathbf{R}_{n,p}'^{-1}\mathbf{a}_{\tau}(\tau_p)}.
  \label{eq:IAA_solu1}
\end{equation}

According to equation \eqref{eq:R_interfere} and the matrix inversion lemma, equation \eqref{eq:IAA_solu1} can be simplified as
\begin{equation}
  \hat{\beta}_{n,p} = \frac{\mathbf{a}_{\tau}^{\mathrm{H}}(\tau_p)\mathbf{R}_{n}^{-1}\mathbf{h}_n'}{\mathbf{a}_{\tau}^{\mathrm{H}}(\tau_p)\mathbf{R}_{n}^{-1}\mathbf{a}_{\tau}(\tau_p)}.
  \label{eq:IAA_solu2}
\end{equation}

Since \(\mathbf{R}_n\) in equation \eqref{eq:IAA_solu2} is unknown, IAA solves the problem by estimating \(\boldsymbol{\beta}_n\) and \(\mathbf{R}_n\) alternatively and iteratively using equation \eqref{eq:iaa_iter1} to equation \eqref{eq:iaa_iter3}:
\begin{align}
  \hat{\beta}_{n,p} &= \frac{\mathbf{a}_{\tau}^{\mathrm{H}}(\tau_p)\hat{\mathbf{R}}_{n}^{-1}\mathbf{h}_n'}{\mathbf{a}_{\tau}^{\mathrm{H}}(\tau_p)\hat{\mathbf{R}}_{n}^{-1}\mathbf{a}_{\tau}(\tau_p)},\quad p = 1,\dots, P, \label{eq:iaa_iter1}\\
  \hat{\mathbf{P}}_n &= \text{diag}\left(\left[\left|\hat{\beta}_{n,1}\right|^2, \left|\hat{\beta}_{n,2}\right|^2, \dots, \left|\hat{\beta}_{n,P}\right|^2\right]^{\mathrm{T}}\right), \label{eq:iaa_iter2}\\
  \hat{\mathbf{R}}_n &= \mathbf{A}_{\tau}\hat{\mathbf{P}}_n\mathbf{A}_{\tau}^{\mathrm{H}}.\label{eq:iaa_iter3}
\end{align}

The initialization can be performed via the classical periodogram method \cite{kay1981_SpectrumAnalysisMo}. Usually, IAA method converges in \(10-15\) iterations. After convergence, the TOA spectrum estimates \(\hat{\boldsymbol{\beta}}_n, n = 1,\dots,N\) are obtained. According to Fig. \ref{fig:proposed_method_flowchart}, amplitudes of multichannel TOA spectra are averaged
at each TOA grid as follows:
\begin{equation}
  \bar{\boldsymbol{\beta}} = \frac{1}{N}\sum_{n=1}^N\left|\hat{\boldsymbol{\beta}}_n\right|.
\end{equation}

A detection algorithm is then applied to vector \(\bar{\boldsymbol{\beta}}\) to retrieve significant path components, followed with a LOS identification module to discriminate the desired LOS component from those path components. Since signal detection and LOS path identification are not the main concern of this paper, we will not discuss their details.
Denoting the TOA of the LOS path and the corresponding TOA index as \(\hat{\tau}_{\text{LOS}}\) and \(p_{\text{LOS}}\), respectively, then the LOS components from all channels can be collected in a vector \(\mathbf{b}_{\text{LOS}} = \left[\hat{\beta}_{1,p_{\text{LOS}}}, \hat{\beta}_{2,p_{\text{LOS}}},\dots,\hat{\beta}_{N,p_{\text{LOS}}}\right]^{\mathrm{T}}\).

The last step is to estimate the DOA of the LOS path from the vector \(\mathbf{b}_{\text{LOS}}\). Since the LOS and NLOS components have been separated in the TOA domain, the proposed scheme uses the naive CBF to estimate the LOS DOA. As stated in Section \ref{sec:proposed-method}, to offset the direction-dependent antenna errors, the pre-calibrated spatial steering-vector function \(\hat{\mathbf{a}}'(\theta)\) is used instead of the ideal one. 
Therefore, the LOS DOA is obtained by
\begin{equation}
  \hat{\theta}_{\text{LOS}} = \arg\max_{\theta}\left[\hat{\mathbf{a}}_{\theta}'(\theta)\right]^{\mathrm{H}}\mathbf{b}_{\text{LOS}}.
  \label{eq:target_fun_DOA}
\end{equation}
Equation \eqref{eq:target_fun_DOA} is solved by an one-dimensional searching over a pre-defined uniform grid set \(\{\theta_q\}_{q=1}^Q\) which covers the DOA span of \(\mathbb{U}\).

Afterward, one can utilize a positioning and tracking framework, such as that based on the extended Kalman filter (EKF), the unscented Kalman filter (UKF), or the particle filter (PF) \cite{arulampalam2002_TutorialParticleFi}, to exploit the estimated LOS DOA \(\hat{\theta}_{\text{LOS}}\) and LOS TOA \(\hat{\tau}_{\text{LOS}}\) for positioning the target UE in the 2-D plane.

\subsection{Complexity Analysis}
\label{sec:complexity_analysis}

In this section, the computational complexity and storage requirements of the proposed scheme are analyzed and compared with JADE algorithms presented in the literature \cite{vanderveen1997_JointAngleDelay, kotaru2015_SpotFiDecimeterLeva, bazzi2016_SpatiofrequentialSm, li2021_DecimeterLevelIndo, bazzi2016_SingleSnapshotJoin, shamaei2021_JointTOADOAa}. Conventional computational implementations are considered throughout the complexity analysis.

The computational complexity is measured in terms of the number of complex multiplications. According to the flowchart illustrated in Fig. \ref{fig:proposed_method_flowchart}, the calibration of the antenna errors is performed in offline-stage, adding no extra computational burdens to the real-time JADE. Therefore, its complexity is not analyzed. The calibration of channel errors expressed by equation \eqref{eq:channel_calibration} can be implemented by multiplying the receiving data with the pre-stored reciprocal of the RF channel responses \(1/\left[\boldsymbol{\Gamma}\right]_{m,n},m=1,\dots,M,n=1,\dots,N\). The resultant complexity is \(\mathcal{O}(MN)\). In the proposed scheme, CBF is applied only to the extracted LOS path component, which has the complexity of \(\mathcal{O}(NQ)\). Hence, the TOA spectrum estimation for each channel dominates the computational burden of the proposed scheme.

According to equation \eqref{eq:iaa_iter1} to equation \eqref{eq:iaa_iter3}, in each iteration, the most computational intensive steps of the IAA-based TOA spectrum estimator are the calculation of the covariance matrix \(\hat{\mathbf{R}}_n\) and its inverse \(\hat{\mathbf{R}}_n^{-1}\), and the grid-by-grid searching of the IAA spectral value \(\beta_{n,p}\). Their complexities are \(\mathcal{O}(PM^2), \mathcal{O}(M^3)\), and \(\mathcal{O}(PM^2)\), respectively. Denoting the iteration steps of the IAA method as \(n_i\), then its total computational complexity is \(\mathcal{O}\left(n_i\left(PM^2 + M^3\right)\right)\). As stated before, the IAA method converges fast, thus the iteration number \(n_i\) is small and is independent on the problem scale. Therefore, the complexity of the IAA-based TOA spectrum estimation for all receiving channels is \(\mathcal{O}(NPM^2+NM^3)\) and the overall computational complexity for the proposed method is \(\mathcal{O}\left(NPM^2+NM^3+MN+NQ\right)\), which is also on the order of \(\mathcal{O}\left(NPM^2+NM^3\right)\). Since the number of TOA search grids \(P\) is usually much larger than the subcarrier number \(M\), the total computational complexity can also be deducted to \(\mathcal{O}(NPM^2)\).

For the IAA-based TOA spectrum estimation, as the proposed method estimates TOA spectrum for each channel sequentially, only the intermediate results for one-channel need to be taken into account for storage requirement analysis. 
It can be inferred from equation \eqref{eq:iaa_iter1} to equation \eqref{eq:iaa_iter3} that matrix \(\mathbf{A}_{\tau}\) and \(\mathbf{R}_n\) need to be stored for the IAA method. The resultant space complexity is \(\mathcal{O}(MP + M^2)\). Besides, memory requirement for storing \(N\)-channel TOA spectrum is \(\mathcal{O}(NP)\). Since the actual spatial structure of the antenna array is modeled as a function \(\mathbf{a}_\theta'(\theta)\),
only a few coefficients for that function, not the entire manifold over the angular set \(\{\theta_q\}_{q=1}^Q\), need to be stored. Therefore, the total space complexity for the proposed method is \(\mathcal{O}(MP+M^2+NP)\).

TABLE \ref{tab:complexity} also lists the computational complexities and storage requirements of two representative JADE methods in the literature for comparison. The first one is the 2-D MUSIC-based JADE method \cite{vanderveen1997_JointAngleDelay, kotaru2015_SpotFiDecimeterLeva, bazzi2016_SpatiofrequentialSm}. Although \cite{kotaru2015_SpotFiDecimeterLeva} and \cite{bazzi2016_SpatiofrequentialSm} improve the original 2-D MUSIC method proposed in \cite{vanderveen1997_JointAngleDelay} by employing the spatial-frequential smoothing technique, the dimension of the smoothed matrix is still linear related to that of the original matrix. Therefore, they have the same computational complexity and space requirement. Their computational complexity is dominated by the EVD of a \(MN\times{}MN\) covariance matrix and a 2-D parameter search, whose complexities are \(\mathcal{O}\left(M^3N^3\right)\) and \(\mathcal{O}\left(PQM^2N^2\right)\), respectively. They all need to store the spatial-frequential signature matrix and the covariance matrix, whose storage requirements are \(\mathcal{O}\left(PQMN\right)\) and \(\mathcal{O}\left(M^2N^2\right)\), respectively. The other one is the search-free 2-D matrix-pencil-based JADE method \cite{li2021_DecimeterLevelIndo, bazzi2016_SingleSnapshotJoin, shamaei2021_JointTOADOAa}. Its most computational intensive step is the SVD of the constructed enhanced-matrix, whose dimension is determined by the frequency-domain and space-domain pencil parameters which are linear related to the subcarrier number \(M\) and array element number \(N\), respectively. Therefore, its computational complexity and space requirement can be denoted as \(\mathcal{O}\left(M^3N^3\right)\) and \(\mathcal{O}\left(M^2N^2\right)\), respectively.

\begin{table}[htbp]
  \caption{Summary of computational complexity and storage requirement}
  \begin{center}
    \begin{tabular}{ccc}
      \toprule
      \bf{JADE algorithm} & \bf{Time complexity} & \bf{Space complexity} \\
      \midrule
      \begin{tabular}[c]{@{}c@{}}MUSIC- \\ based\cite{vanderveen1997_JointAngleDelay, kotaru2015_SpotFiDecimeterLeva, bazzi2016_SpatiofrequentialSm}\end{tabular} & \(\mathcal{O}\left(M^3N^3 + PQM^2N^2\right)\) & \(\mathcal{O}\left(PQMN + M^2N^2\right)\)\\
      \begin{tabular}{@{}c@{}}Matrix-pencil- \\ based\cite{li2021_DecimeterLevelIndo, bazzi2016_SingleSnapshotJoin, shamaei2021_JointTOADOAa}\end{tabular} & \(\mathcal{O}\left(M^3N^3\right)\) & \(\mathcal{O}\left(M^2N^2\right)\)\\
      Proposed & \(\mathcal{O}\left(NPM^2+NM^3\right)\) & \(\mathcal{O}(MP+M^2+NP)\) \\
      \bottomrule
    \end{tabular}
    \label{tab:complexity}
  \end{center}
\end{table}

As shown in TABLE \ref{tab:complexity}, the proposed method has a much lower computational complexity and storage requirement compared with these popular 2-D subspace-based JADE methods.
For CS-based JADE methods \cite{gong2017_RobustIndoorWirelea, zhang2019_BreathTrackTracking, fan2021_LocalizationBasedI, liu2021_IndoorPassiveLocala}, their complexities are dependent on the underlying sparse recovery methods. Usually, they have to deal with a huge-scale 2-D overcomplete dictionary, and incorporate a solver for a large-dimensional convex optimization problem \cite{gong2017_RobustIndoorWirelea, zhang2019_BreathTrackTracking} or an iterative procedure with complicated matrix operations at each iteration \cite{fan2021_LocalizationBasedI, liu2021_IndoorPassiveLocala}. Therefore, they also exhibit much higher computational and storage burden than the proposed method.


\section{Enhancements for Efficient Implementation}
\label{sec:impl-issu}

As analyzed in Section \ref{sec:complexity_analysis} and illustrated in TABLE \ref{tab:complexity}, although the computation and storage budget of the proposed JADE scheme have been largely reduced compared with those 2-D joint estimation methods in the literature, challenges for its real-time implementation still remain, especially in time-sensitive applications. In this section, enhancements to the proposed scheme are proposed to further reduce complexity. Concretely, a CFR pre-processing module is prepended to facilitate the following estimation procedures, and the IAA-based TOA spectrum estimator, which dominates the computational complexity, is accelerated by employing FFTs. 


\subsection{Pre-Processing: CFR Denoising}
\label{sec:cfr-denoise}


As shown in TABLE \ref{tab:complexity}, the computational complexity of the proposed method increases with the number of subcarriers \(M\) in an increasing rate larger than the quadratic function. On the other hand, the demands of obtaining a finer characterization of the wireless channel and a better resolution for multipath components prompt the positioning system to use a large-bandwidth SRS with a plentiful amount of subcarriers, which gives rise to a large-dimensional CFR matrix.
Therefore, applying the proposed JADE method to this CFR matrix directly will occupy substantial computing resources and lead to an unacceptable processing latency. 

\begin{figure}[htb]
  \centering
  \includegraphics[width=0.48\textwidth]{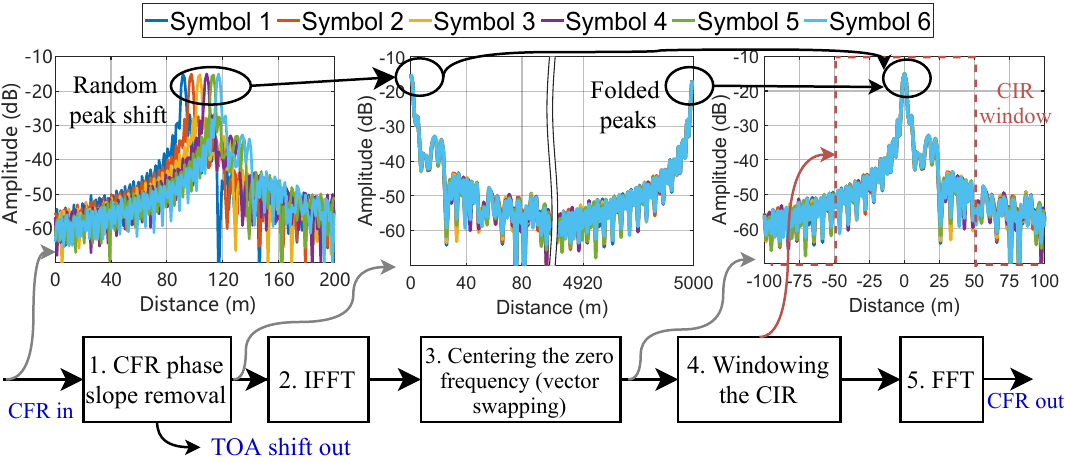}
  \caption{Flowchart of the proposed pre-processing scheme.}
  \label{fig:pre_processing_scheme}
\end{figure}

However, if the wireless channel is investigated in the perspective of the channel impulse response (CIR), i.e. the time-domain counterpart of the CFR, the delay taps of the CIRs that exceed the maximum delay covered by the gNB make no contribution to JADE and bring extra noises into the CFR \cite{chen2020_AoAawareProbabilist}. For a picocell gNB with a coverage of several tens of meters, the valid delay taps occupy only a small proportion of the entire CIR. For example, assuming that the SRS is configured in accordance with TABLE \ref{tab:srs_parameters}, since the comb-two structure is used, the spacing between two adjacent SRS subcarriers is \(60\;\mathrm{kHz}\), which corresponds to a maximum unambiguous delay of \(16.67\;\mathrm{\mu s}\) in the CIR. For a gNB coverage of \(100\;\mathrm{m}\), the first \(2\%\) delay taps of the CIR already preserve all the information of the LOS and multipath components. Hence, by truncating the CIR and re-transforming it to the frequency domain, the CFR can be recovered with noise suppressed and dimension reduced. Based on this observation, a pre-processing scheme for the purposes of denoising and dimension-reducing is proposed, whose flowchart is shown in Fig. \ref{fig:pre_processing_scheme}.

Fig. \ref{fig:pre_processing_scheme} also considers the random TOA offsets between consecutive SRS symbols caused by the imperfect synchronization between the UE and the gNB \cite{goodarzi2021_Synchronization5GNa}. In the presence of this additional offset, the desired CIR taps with propagation information are randomly shifted from the zero point, and it is improper to apply a low-pass filter with an identical cutoff delay of \(\tau_{\mathrm{max}}\) to the CIRs estimated from different SRS symbols. The proposed pre-processing scheme deals with this realistic issue by adding a CFR phase slope removing module (Step 1)
in the front of the denoising modules (Step 2 to 5). The phase slope removal approach similar to those proposed in \cite{sen2012_YouAreFacing} and \cite{kotaru2015_SpotFiDecimeterLeva} for the sanitization of Wi-Fi sensed channel state information (CSI) can be used here. After this phase slope removal operation, the main peak of the CIR is nearly centered at the zero delay.

In the flowchart of Fig. \ref{fig:pre_processing_scheme}, the CIR plots at crucial steps are illustrated for consecutive SRS symbols transmitted by a non-moving user terminal and collected by a 5G picocell gNB in an indoor environment. Parameters of the SRS conform to TABLE \ref{tab:srs_parameters}. Point for inverse FFT (IFFT) in Step 2 is set to be \(1632\) and the delay window in Step 4 is selected to be \(\left[-166.67\;\mathrm{ns}, +166.67\;\mathrm{ns}\right]\), which corresponds to \(\left[-50\;\mathrm{m}, +50\;\mathrm{m}\right]\) in distance. Then in this example, only \(41\) points in the CIR fall in this window, and the FFT point in Step 5 can be set as \(64\) accordingly.

\begin{figure}[htb]
  \centering
  \subfloat[Before pre-processing.]{\includegraphics[width=0.23\textwidth]{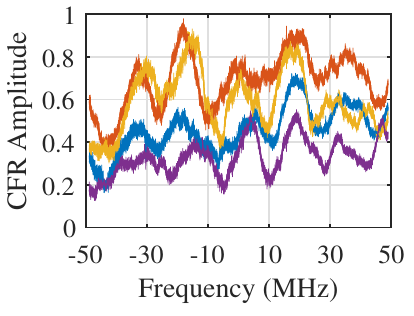}} 
  \subfloat[After pre-processing.]{\includegraphics[width=0.23\textwidth]{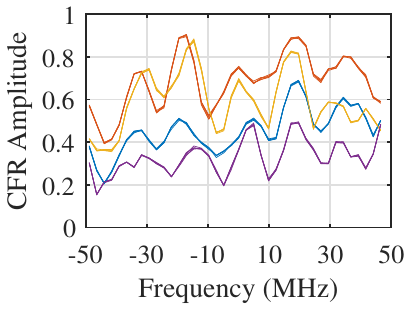}}\\
  \subfloat{\includegraphics[width=0.4\textwidth]{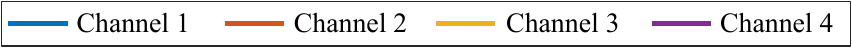}}
  \caption{Impact of pre-processing on CFR amplitude.}
  \label{fig:csi_denoise}
\end{figure}

The effects of the proposed pre-processing scheme are demonstrated from the aspects of denoising and complexity reduction. First, the denoising effect is illustrated by examining the CFR amplitudes for these six consecutive SRS symbols shown in the example of Fig. \ref{fig:pre_processing_scheme}. The IFFT point, delay window, and FFT point are also set to be \(1632, \left[-166.67\;\mathrm{ns}, +166.67\;\mathrm{ns}\right]\), and \(64\), respectively. CFR amplitudes before and after this pre-processing are respectively shown in Fig. \ref{fig:csi_denoise}(a) and Fig. \ref{fig:csi_denoise}(b). They clearly show that the output CFR is a smoothed and noise reduced version of the input one. 
Second, denoting the subcarrier number of the SRS after pre-processing as \(\tilde{M}\), then the computational complexity of this pre-processing module is nearly \(\mathcal{O}(M\log_2M+\tilde{M}\log_2\tilde{M})\). According to the analysis in Section \ref{sec:complexity_analysis}, the complexity of the proposed JADE method based on the dimension-reduced CFR matrix is \(\mathcal{O}\left(NP\tilde{M}^2+N\tilde{M}^3\right)\). Similar analysis can be applied to its storage requirement. Since \(\tilde{M}\ll M\), the overall time and space complexities can be largely reduced when prepending the proposed pre-processing module to the front of the JADE modules.



\subsection{TOA Spectrum Estimation: Accelerating Spectrum Computing by FFTs}
\label{sec:fastIAA}
In this section, based on the evenly-spaced comb pattern of the SRS \cite{3gpp.38.211} and the resulting Vandermonde structure of the delay signature matrix, as revealed by equation (\ref{eq:a_tau}), the computing of the TOA spectrum
is accelerated by employing several FFT operations. Concretely, suppose the grid set \(\{\tau_p\}_{p=1}^P\) for TOA spectrum searching spans the whole unambiguous delay range of \(\left[0, (\Delta f)^{-1}\right]\), then according to equation (\ref{eq:a_tau}),
the delay signature matrix \(\mathbf{A}_\tau\) is composed of the first \(M\) rows of the \(P\)-point discrete Fourier transform base matrix. Therefore, \(\mathbf{A}_\tau\)-involved matrix multiplications can be computed via FFTs.
Note that whether FFT or IFFT is used depends on the sign of the complex exponential for the delay signature matrix, and we use FFT uniformly here.

First, denoting
\begin{align}
  \mathbf{Q} &= \hat{\mathbf{R}}_n^{-1}, \\
  \boldsymbol{\iota} &= \mathbf{Q}\mathbf{h}_n,
\end{align}
then the numerators of equation (\ref{eq:iaa_iter1}) for all \(P\) searching grids can be obtained by applying a single FFT on vector \(\boldsymbol{\iota}\) and gathered in a vector \(\boldsymbol{\rho}\).
That is, \(\boldsymbol{\rho} = \mathsf{FFT}[\boldsymbol{\iota}, P]\).

Secondly, since the denominator of equation (\ref{eq:iaa_iter1}) is identical to that of the classical Capon spectral estimator, then according to the proof in \cite{Li1998_Caponestimationcovariance}, its values for all the \(P\) searching grids can also be computed in a batch using a FFT. Specifically, collecting these values in a vector \(\boldsymbol{\xi}\), it can be computed by
\begin{equation}
\boldsymbol{\xi} = \mathsf{FFT}[\mathbf{u}, P],
\end{equation}
where the vector \(\mathbf{u}\in\mathbb{C}^{P\times 1}\) is:
\begin{equation}
  \mathbf{u}=\left[u_{0}, \dots, u_{M-1}, \mathbf{0}_{1\times(P+1-2M)}, u_{1-M},\dots, u_{-1}\right]^{\mathrm{T}},
\end{equation}
in which for \(m = -(M-1),\dots, M-1\), \(u_m\) is the summation of the \(m\)-th diagonal of the matrix \(\mathbf{Q}\):
\begin{equation}
  u_{m}=\sum_{i=\max (0, m)}^{\min (M-1+m, M-1)} \left[\mathbf{Q}\right]_{i+1, i-m+1}.
\end{equation}

Thirdly, since \(\mathbf{A}_\tau\) is Vandermonde and \(\hat{\mathbf{P}}_n\) is a diagonal matrix, \(\hat{\mathbf{R}}_n\) represented by equation (\ref{eq:iaa_iter3}) is both Toeplitz and Hermitian \cite{backstrom2013_VandermondeFactoriz}. Therefore, it is fully determined by its first row, which can be computed by another FFT on the diagonal elements of \(\hat{\mathbf{P}}_n\) \cite{Xue2011_IAASpectralEstimation}. Specifically, collecting elements of the first row of \(\hat{\mathbf{R}}_n\) in a vector \(\mathbf{r}\in\mathbb{C}^{M\times1}\), then equation (\ref{eq:iaa_iter3}) can be calculated as:
\begin{align}
  \mathbf{r}' &= \mathsf{FFT}\left(\left[\left|\hat{\beta}_{n,1}\right|^2, \left|\hat{\beta}_{n,2}\right|^2, \dots, \left|\hat{\beta}_{n,P}\right|^2\right]^{\mathrm{T}}, P\right), \\
  \mathbf{r} &= \left[\mathbf{r}'\right]_{1:M}, \\
  \hat{\mathbf{R}}_n &= \mathsf{Toeplitz}\left(\mathbf{r}\right).
\end{align}


The IAA method implemented with these three FFTs is denoted as the FFT-IAA method herein. According to the analysis above, by employing FFT-IAA, the total computational complexity of the proposed method can be reduced from \(\mathcal{O}\left(NPM^2 + NM^3\right)\) to \(\mathcal{O}\left(NP\log_2P + NM^3\right)\). Additionally, since the FFT-IAA method does not need to store the delay signature matrix, its storage requirement is also reduced from \(\mathcal{O}\left(MP + M^2 + NP\right)\) to \(\mathcal{O}\left(M^2 + NP\right)\). Since for 5G picocell gNBs, \(P\gg M\gg N\), much lower time and space complexities are achieved. Actually, the proposed scheme presented in Section \ref{sec:proposed-method} combined with the CFR denoising module and this FFT-IAA method has already been implemented and deployed in the positioning server for field test, as will be presented in Section \ref{sec:indoor-exp}.







\section{Performance Evaluation}
\label{sec:perf-eval}

In this section, the performance of the proposed method is evaluated by the following series of elaborately designed experiments:
\begin{enumerate}
\item First, the effectiveness of the calibration part of the proposed scheme is validated by multipath-free signals collected in an anechoic chamber; 
\item Next, the resolution and accuracy of the proposed method in a dense multipath environment are demonstrated by simulated multipath CFR data conforming to the signal model represented by equation (\ref{eq:revised_H});
\item After that, the performance of the proposed method in typical indoor environments is further evaluated by simulated 5G channel data whose parameters conform to the standard 5G indoor channel models \cite{3gpp.38.901} and by field test data collected in an underground parking lot, respectively;
\item Lastly, we conclude the performance evaluations by examining the running time of the proposed method.
\end{enumerate}

Throughout these experiments, the proposed method is compared with existing JADE methods, including those based on the classical periodogram method \cite{kay1981_SpectrumAnalysisMo}, the smoothed-MUSIC method \cite{kotaru2015_SpotFiDecimeterLeva, bazzi2016_SpatiofrequentialSm}, and the matrix-pencil method \cite{li2021_DecimeterLevelIndo, bazzi2016_SingleSnapshotJoin, shamaei2021_JointTOADOAa}. The pre-processing scheme as shown in Fig. \ref{fig:pre_processing_scheme} is applied before all these JADE methods to denoise and to reduce complexity. The counts of IFFT points in Step 2 and FFT points in Step 5 of this pre-processing scheme are chosen to be \(2048\) and \(64\), respectively. For the smoothed-MUSIC-based JADE method, the smoothing orders in frequency-domain and space-domain are set to be \(6\) and \(2\), respectively; while for the matrix-pencil-based JADE method, the pencil parameters in these two domains are set to be \(35\) and \(3\), respectively.

During all the experiments, a picocell gNB configuration is considered. When considering the estimation of the DOA and TOA of the uplink SRS impinged at a specific TRP, the TRP is assumed to locate at \([0,0]\;\mathrm{m}\) in the XOY plane, and the UE is always in the sector area with the azimuth angle spanning from \(-60^\circ\) to \(60^\circ\) and with the maximum range of \(50~\mathrm{m}\).
Then for search-based JADE methods, such as the periodogram-based, the smoothed-MUSIC-based, and the proposed, the DOA and TOA are searched in the range of \([-60^\circ, 60^\circ]\) and \([0, 50]\;\mathrm{m}\) with a grid size of \(0.2^\circ\) and \(0.2~\mathrm{m}\), respectively.

The SRS parameters are all configured according to TABLE \ref{tab:srs_parameters} for experiments in this section. A batch of antenna arrays and 5G picocell RRUs identical to that shown in Fig. \ref{fig:rru_antenna} is fabricated and established for experiments in the anechoic chamber and indoor environment. Their calibration coefficients are pre-measured using the approaches introduced in Section \ref{sec:array-modeling-error}. For numerical simulation experiments, an identical four-element ULA with a half-wavelength spacing is used. In addition, the calibration coefficients of a randomly selected antenna array are used to generate received signals with direction-dependent antenna errors, while the RF channel errors are assumed to be perfectly compensated.

Besides, the time synchronization error between the UE and the gNB induces a severe offset in the TOA estimates, which is hard to be separated from the absolute propagation delays in free-space \cite{goodarzi2021_Synchronization5GNa}. Hence, for simulations, we assume perfect timing, while for field test in the indoor environment, differentials of the TOA estimates from two TRPs, i.e. the time differences of arrival (TDOAs), are evaluated instead.


\subsection{Field Test in an Anechoic Chamber}
\label{sec:anechoic-exp}
The experimental setup in the anechoic chamber for performance evaluation is illustrated by Approach (b) of Fig. \ref{fig:chamber_setups}. Different from Approach (a) which measures the array response using dedicated instruments,
Approach (b) establishes the whole 5G infrastructure in this anechoic chamber and employs the 5G UE and gNB for signal transmitting and receive signal processing, respectively. 
Concretely, similar to Approach (a), the receiving ULA is still rotated from \(-60^\circ\) to \(+60^\circ\) by an angular interval of \(5^\circ\) during the experiments. The difference is that, at each angle, \(500\) SRS symbols transmitted by the 5G UE are conducted to the horn antenna and the sounded CFRs are collected by the gNB.

\begin{figure}[htb]
  \centering
  \includegraphics{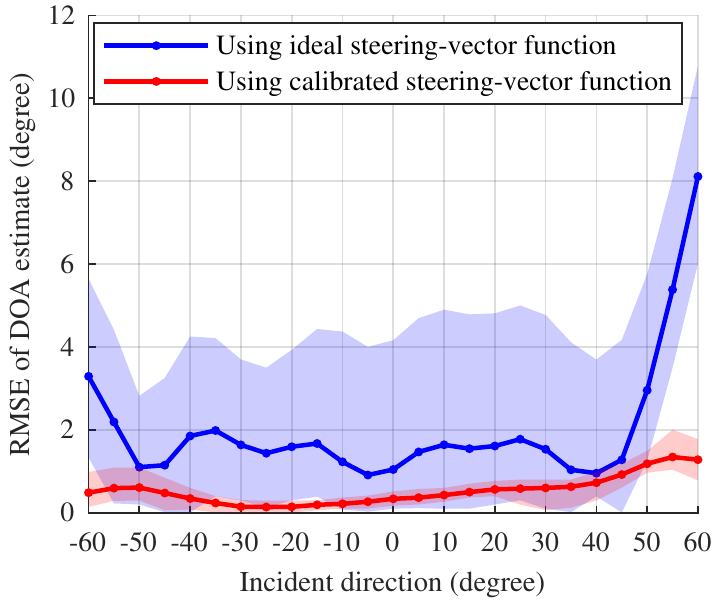}
  \caption{RMSE statistics of DOA estimates obtained by \(12\) antennas in an anechoic chamber when incident direction varies from \(-60^\circ\) to \(+60^\circ\).}
  \label{fig:rmse_chamber_compare}
\end{figure}

To illustrate the repercussions caused by the array modeling errors and demonstrate the effectiveness of the calibration module in the proposed JADE scheme, the DOA estimation errors of the proposed method with the ideal and calibrated steering-vector functions are compared in Fig. \ref{fig:rmse_chamber_compare}.

Fig. \ref{fig:rmse_chamber_compare} demonstrates the root mean square errors (RMSEs) of the DOA estimates obtained by experiments with \(12\) different antennas, each of which is calculated based on the DOA measures from consecutive SRS symbols collected by that antenna as follows:
\begin{equation}
\mathrm{RMSE} = \sqrt{\frac{1}{L}\sum_{l=1}^L\left(\hat{\theta}_l - \theta_{\mathrm{truth}}\right)^2},
\end{equation}
where \(\hat{\theta}_l\) represents the estimate for \(l-\mathrm{th}\) SRS, \(\theta_{\mathrm{truth}}\) represents the DOA ground truth, and \(L\) is the count of measurements, which equals to \(500\) for Fig. \ref{fig:rmse_chamber_compare}.

In Fig. \ref{fig:rmse_chamber_compare}, at each angle, the solid dot shows the average value of RMSEs from these \(12\) antennas, and the lower and upper bounds of the shadowed area mark the minimum and maximum RMSE values, respectively. Fig. \ref{fig:rmse_chamber_compare} indicates that, estimating DOA using the calibrated steering-vector function counteracts the direction-dependent antenna modeling errors dramatically, especially when the signal impinges the receiving array from directions larger than \(\pm45^\circ\). For example, the averaged DOA estimation error can be reduced from \(8.11^\circ\) to \(1.28^\circ\) at the incident direction of \(+60^\circ\).
This phenomenon is attributable to the divergence of the antenna phase errors in large directions, which has been evidently shown in Fig. \ref{fig:phase_differences_angles}.

\subsection{Numerical Experiments Using Simulated Multipath Data}
\label{sec:simu-multipath}

CFR data with direction-dependent array modeling errors is simulated based on equation (\ref{eq:revised_H}). During experiments, the multipath number varies from three to five, which is representative for typical indoor environments. The signal-to-noise-ratio (SNR) of the received signal varies from \(-10\) decibels (dB) to \(10\;\mathrm{dB}\). \(500\) Monte Carlo trials are conducted for each scenario with a specific multipath number and a specific SNR. The fading coefficients for all multipath components are assumed to be identical. In each Monte Carlo trial, the DOA and TOA of each multipath component and the realization of the wideband multichannel noise component are generated randomly. Among them, the DOAs and TOAs are generated according to the uniform distributions over \(-60^\circ\) to \(+60^\circ\) and over \(0\) to \(166.67\;\mathrm{ns}\;(50\;\mathrm{m})\), respectively, i.e. \(\theta_k\sim\mathcal{U}\left(-60, +60\right], \tau_k\sim\mathcal{U}\left(0, 166.67\;\mathrm{ns}\right]\), and the real and imaginary parts of the noise component are identically and independently distributed (i.i.d), each follows a standard Gaussian distribution.

\begin{figure*}[htb]
  \centering
  \subfloat[DOA errors (path number: \(3\), SNR: \(-10\;\mathrm{dB}\)).]{\includegraphics{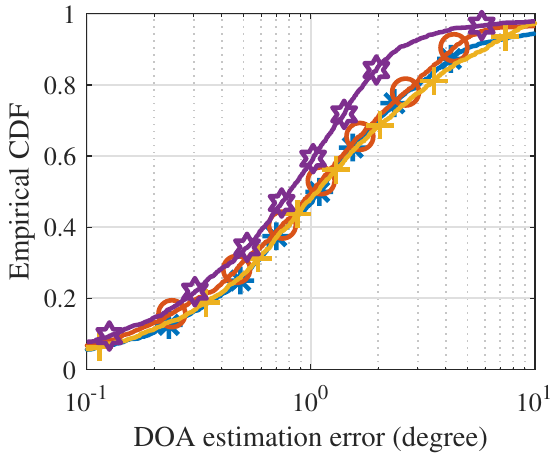}}
  \hfill
  \subfloat[DOA errors (path number: \(4\), SNR: \(-10\;\mathrm{dB}\)).]{\includegraphics{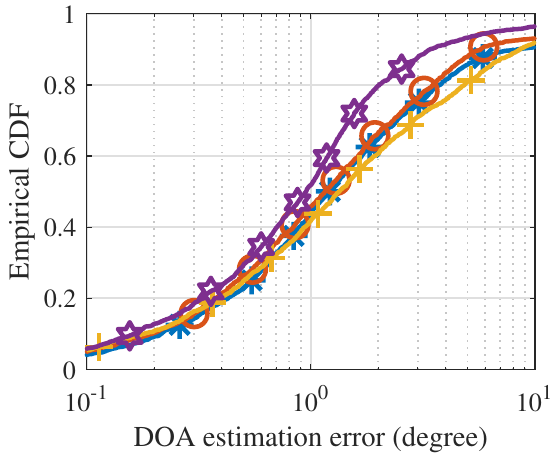}}
  \hfill
  \subfloat[DOA errors (path number: \(5\), SNR: \(-10\;\mathrm{dB}\)).]{\includegraphics{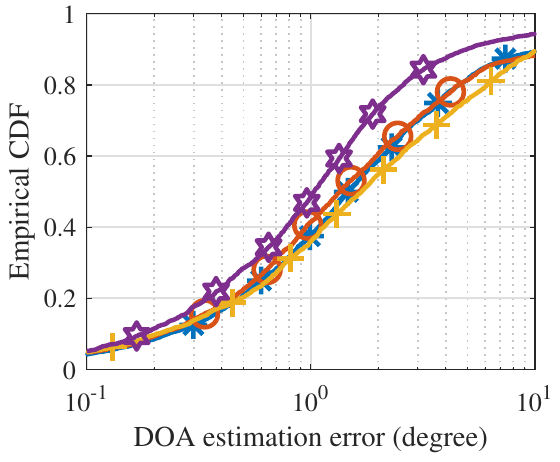}}\\
  \subfloat[TOA errors (path number: \(3\), SNR: \(-10\;\mathrm{dB}\)).]{\includegraphics{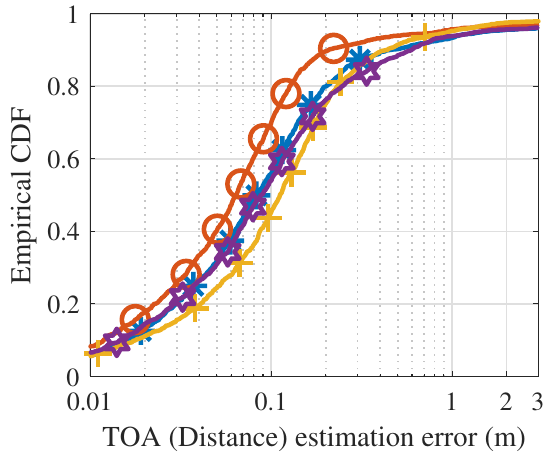}}
  \hfill
  \subfloat[TOA errors (path number: \(4\), SNR: \(-10\;\mathrm{dB}\)).]{\includegraphics{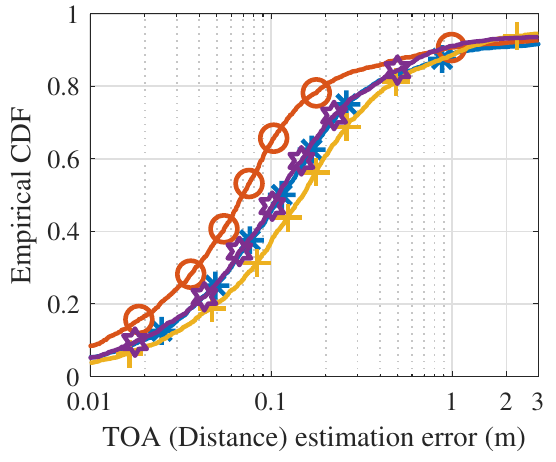}}
  \hfill
  \subfloat[TOA errors (path number: \(5\), SNR: \(-10\;\mathrm{dB}\)).]{\includegraphics{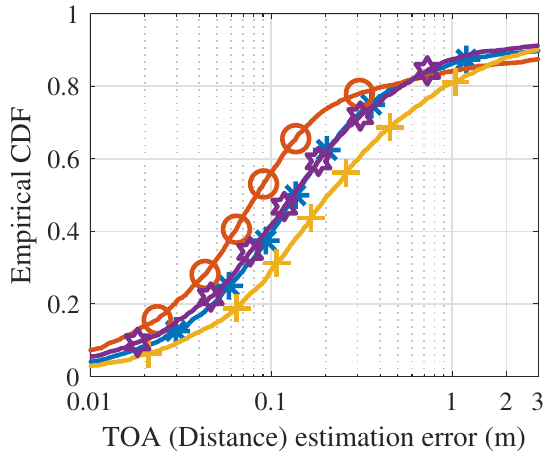}}\\
  \subfloat[Positioning errors (path number: \(3\)).]{\includegraphics{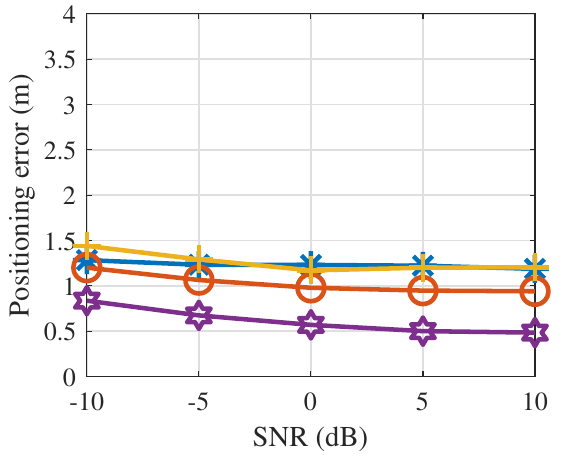}}
  \hfill
  \subfloat[Positioning errors (path number: \(4\)).]{\includegraphics{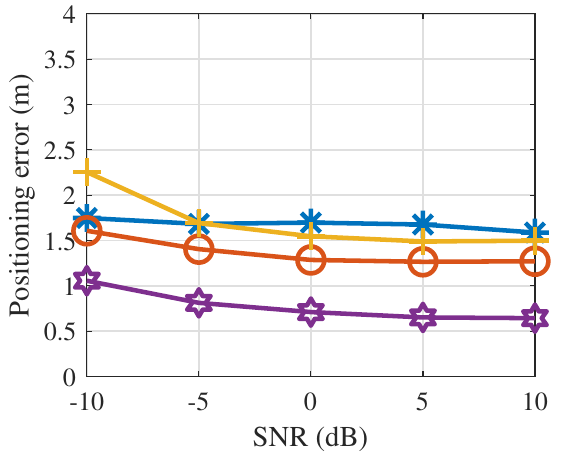}}
  \hfill
  \subfloat[Positioning errors (path number: \(5\)).]{\includegraphics{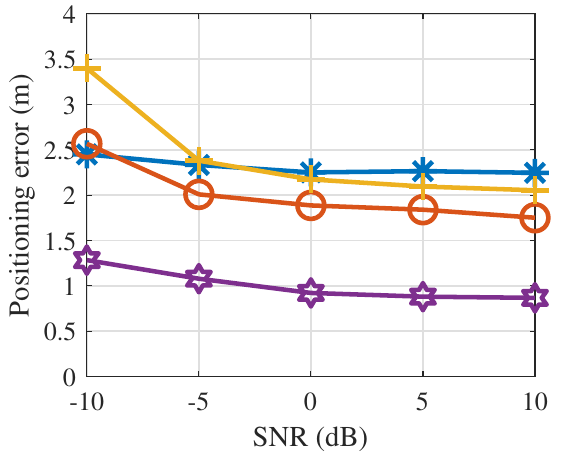}}\\
  \subfloat{\includegraphics{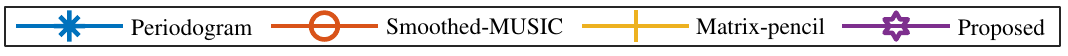}}
  \caption{Performance comparisons of JADE methods by simulated multipath data with path number varies from three to five. (a)-(c) and (d)-(f) show the empirical cumulative distribution function (CDF) curves under the SNR of \(-10\;\mathrm{dB}\) for DOA estimation and TOA (distance) estimation errors, respectively. (g)-(i) show the \(80\)-th percentiles of the positioning errors based on the DOA and TOA estimates when the SNR varies from \(-10\;\mathrm{dB}\) to \(10\;\mathrm{dB}\).}
  \label{fig:result_multipath_simulation}
\end{figure*}

The simulation results are shown in Fig. \ref{fig:result_multipath_simulation}.
Firstly, Fig. \ref{fig:result_multipath_simulation}(a) to \ref{fig:result_multipath_simulation}(c) indicate that, although a relatively naive CBF method is used for DOA estimation in the proposed scheme, its DOA estimation performance is still superior to all the other JADE methods. For example, from Fig. \ref{fig:result_multipath_simulation}(c), when there are five strong multipaths and the SNR is \(-10\;\mathrm{dB}\), the proposed method can achieve a reduction of at least \(44\%\) for the \(80\)-th percentile of the DOA estimation error (from \(4.58^\circ\) to \(2.58^\circ\)).

Secondly, since the proposed method performs TOA spectral analysis for each channel individually and then averages the amplitudes of these TOA spectra, it integrates all these multichannel information non-coherently. On the contrary, the other three JADE methods employ 2-D spectral estimators, which process the multichannel signals coherently. Therefore, it is plausible that these methods outperform the proposed method in the TOA domain. But instead, the 2-D periodogram and 2-D matrix-pencil methods exhibit performance degradation: for the 2-D periodogram, this owes to its limited resolving ability; while for the 2-D matrix-pencil, this attributes to the destruction of the shift-invariance property by the direction-dependent array modeling error. From Fig. \ref{fig:result_multipath_simulation}(f), when there are five strong multipaths and the SNR is \(-10\;\mathrm{dB}\), the \(80\)-th percentile of the TOA estimation error increases only by \(0.3\;\mathrm{ns}\;(0.09\;\mathrm{m})\) for the proposed method when compared with the 2-D smoothed-MUSIC-based JADE method (From \(1.45\;\mathrm{ns}\;(0.435\;\mathrm{m})\) to \(1.75\;\mathrm{ns} \;(0.525\;\mathrm{m})\)).

Lastly, the \(80\)-th percentiles of the positioning errors based on these DOA and TOA estimates are shown in Fig. \ref{fig:result_multipath_simulation}(g) to Fig. \ref{fig:result_multipath_simulation}(i). The location of the UE is derived from the DOA and TOA estimated by a single TRP, and the positioning error is calculated as the Euclidean distance between the derived position and the ground truth. It can be inferred that a much better positioning accuracy can be achieved based on the proposed method than those based on the existing JADE methods. Also according to the simulation, when the path number is no more than \(5\) and the receiving SNR is above \(-10\;\mathrm{dB}\), the single-site positioning error achieved by the proposed method is below \(1.3\;\mathrm{m}\) for \(80\%\) cases.


\subsection{Numerical Experiments Using Simulated 5G Wireless Channel Data}
The performance of the proposed method is then evaluated with realistic wireless channel data generated by the QuaDRiGa channel simulator \cite{jaeckel2014_QuaDRiGa3DMultiCel}. The configurations of \(\mathsf{3GPP\_38.901\_InF\_LOS}\) and \(\mathsf{3GPP\_38.901\_Indoor\_LOS}\) for FR1 band are selected for the simulator, which represent the indoor factory LOS (InF-LOS) and the indoor office LOS (InO-LOS) scenarios, respectively \cite{3gpp.38.901}. With the transmitting power of the UE and the noise figure of the gNB fixed to \(200\;\mathrm{mW}\) and \(5\;\mathrm{dB}\), respectively, the power of the receiving signal and noise component are derived by the QuaDRiGa simulator according to the propagation model.

The default Ricean K-factors for 3GPP 38.901 InF-LOS and InO-LOS channels are both \(7\;\mathrm{dB}\) \cite{3gpp.38.901}. To evaluate the parameter estimation performance when the multipath power varies, the Ricean K-factors for both channels are configured to vary from \(0\;\mathrm{dB}\) to \(7\;\mathrm{dB}\) during the experiments. For each scenario with a specific Ricean K-factor, \(1000\) Monte Carlo trials are conducted, and similar to the approach stated in Section \ref{sec:simu-multipath}, the ground truths of the DOAs and TOAs are randomly generated in each trial.

\begin{figure*}[htb]
  \centering
  \subfloat[DOA estimation errors.]{\includegraphics{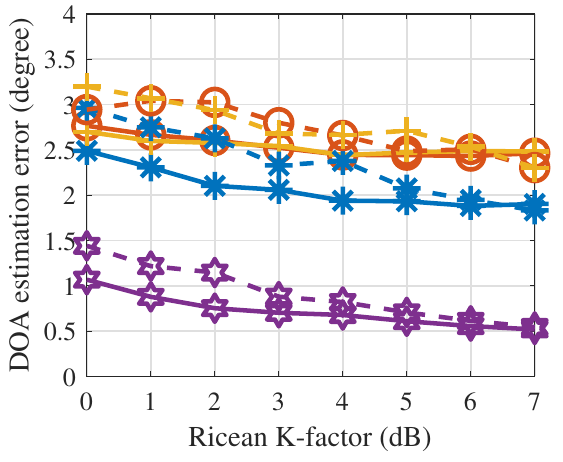}}
  \hfill
  \subfloat[TOA (Distance) estimation errors.]{\includegraphics{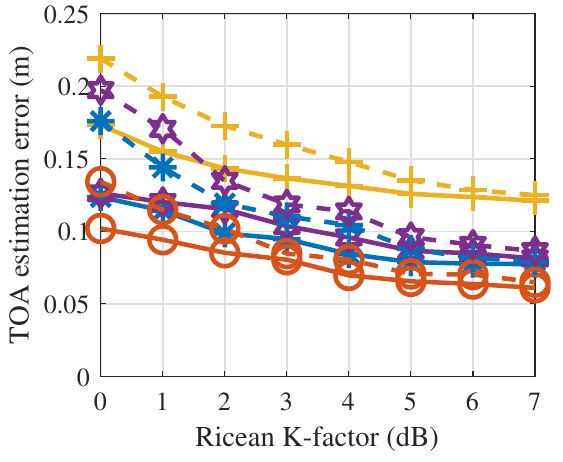}}
  \hfill
  \subfloat[Positioning errors.]{\includegraphics{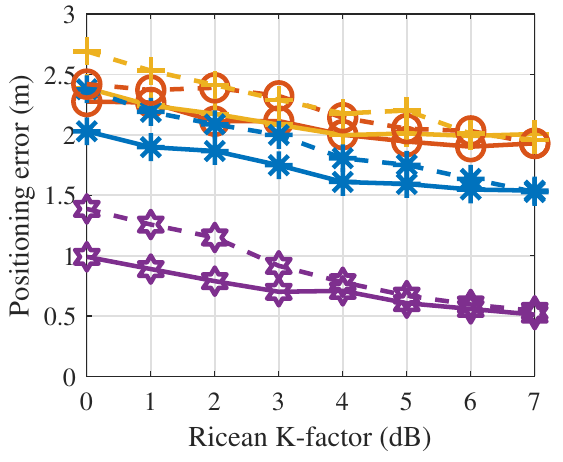}}\\
  \subfloat{\includegraphics{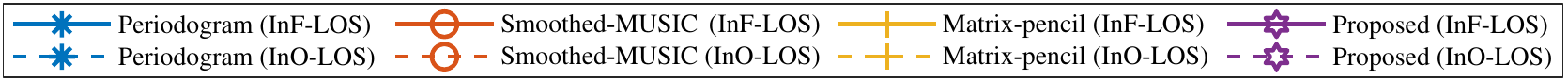}}
  \caption{Performance comparisons of JADE methods by simulated 5G InF-LOS and InO-LOS channel data whose parameters conform to the 3GPP 38.901 report \cite{3gpp.38.901} except their Ricean K-factors, which vary from zero to seven decibels to generate multipath components with different powers. (a), (b), and (c) show the \(80\)-th percentiles for the DOA, TOA (distance) and position estimation errors, respectively.
  }
  \label{fig:result_5G_channel_simulator}
\end{figure*}

The Monte Carlo evaluation results for DOA, TOA and 2-D position estimations are summarized in Fig. \ref{fig:result_5G_channel_simulator}.
Similar to the results presented in Section \ref{sec:simu-multipath}, in realistic indoor environments, the proposed method has a significant improvement for the DOA estimation performance at the cost of a slightly reduced TOA estimation performance. Furthermore, when the DOA and TOA are exploited to locate the UE, the proposed method also achieves the best single-site positioning performance. For example, for the default InF-LOS scenario, a reduction of at least \(72\%\) (from \(1.8^\circ\) to \(0.5^\circ\)) for DOA estimation error is achieved, while the TOA estimation error only increases by \(0.02\;\mathrm{m}\) (from \(0.06\;\mathrm{m}\) to \(0.08\;\mathrm{m}\)) at most, and the total positioning error is also reduced by at least \(67\%\) (from \(1.54\;\mathrm{m}\) to \(0.51\;\mathrm{m}\)).

Further, according to Fig. \ref{fig:result_5G_channel_simulator}(c), when the Ricean K-factor is above \(0\;\mathrm{dB}\), positioning errors of \(0.99\;\mathrm{m}\) and \(1.39\;\mathrm{m}\) can be achieved by the proposed method for the standard 3GPP 38.901 InF-LOS and InO-LOS scenarios, respectively. It can be also inferred from Fig. \ref{fig:result_5G_channel_simulator} that, all algorithms perform better in the InF-LOS scenario than in the InO-LOS scenario. This is attributable to the smaller delay spread parameter of the InO-LOS channel, which generates more densely distributed multipath components \cite{3gpp.38.901}.

\subsection{Field Test in an Indoor Environment}
\label{sec:indoor-exp}

\begin{figure}[htb]
  \centering
  \includegraphics[width=0.45\textwidth]{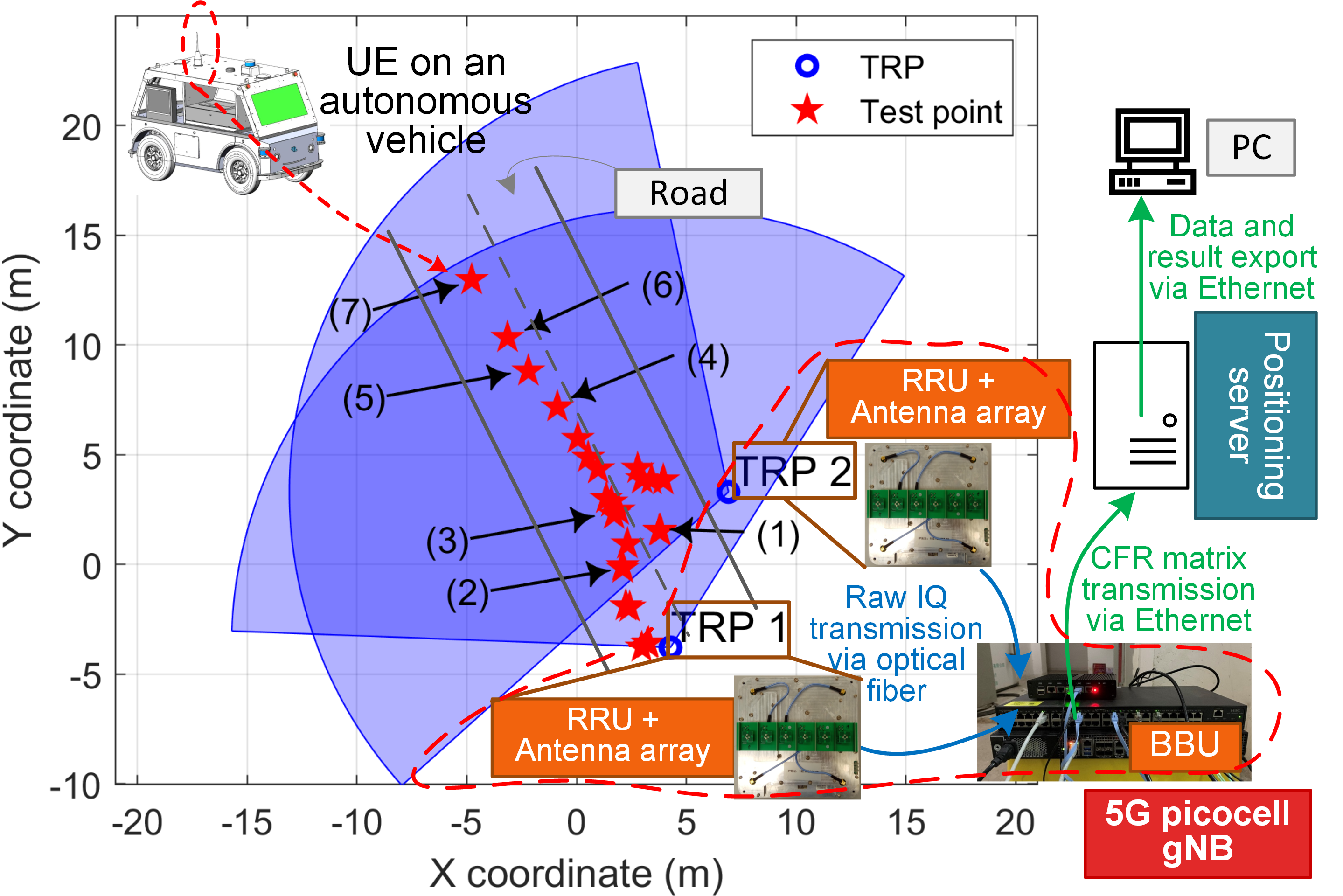}
  \caption{Experiment layout for positioning experiments in an underground parking lot.}
  \label{fig:setup_indoor_experiments}
\end{figure}

To evaluate the performance of the proposed JADE scheme in a real indoor environment, a field test is conducted in an underground parking lot with a commercial 5G picocell gNB established. As illustrated by the experiment layout of Fig. \ref{fig:setup_indoor_experiments}, this gNB contains two TRPs, which are separated by \(7.6\;\mathrm{m}\), and have an overlapping coverage to communicate to the UE simultaneously. Each of these TRPs is composed of the six-element ULA and the picocell RRU as shown in Fig. \ref{fig:rru_antenna}. The 5G UE is deployed on an autonomous vehicle, which can determine its location at the accuracy of several centimeters.

Our proposed JADE method has been implemented and deployed in the positioning server and is able to estimate the DOA and TOA of the uplink SRS in realtime. DOA and TOA estimates and the ground truths given by the autonomous vehicle are exported for offline analysis.
As shown in Fig. \ref{fig:setup_indoor_experiments}, during the experiments, the UE is placed at \(18\) different locations in the lane, and at some locations, the UE is rotated and retested to enlarge the dataset. Therefore, \(30\) cases are tested in total. For each case, the real-time results from \(300\) SRS symbols are collected. Also note that, seven typical test points are marked out in Fig. \ref{fig:setup_indoor_experiments} and results at those individual points will be presented later.

\begin{figure*}[htb]
  \centering
  \subfloat[Empirical CDFs for DOA estimation errors of TRP-1.]{\includegraphics{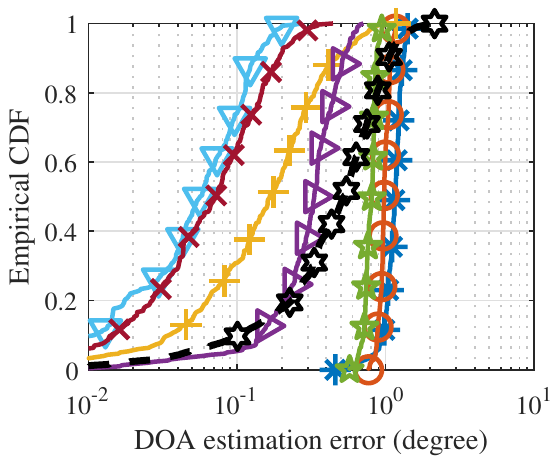}}
  \hfill
  \subfloat[Empirical CDFs for DOA estimation errors of TRP-2.]{\includegraphics{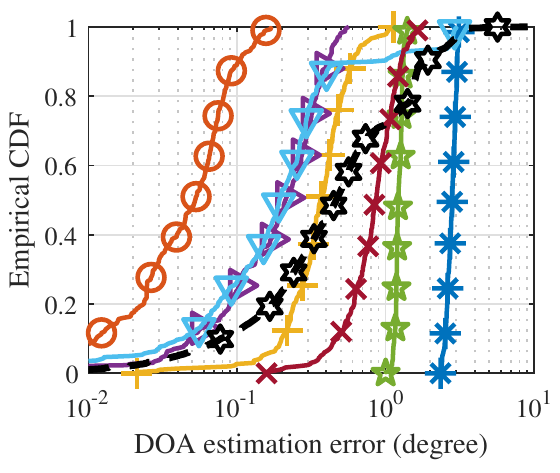}}
  \hfill
  \subfloat[Empirical CDFs for errors of the differentials of TOA estimates at these two TRPs.]{\includegraphics{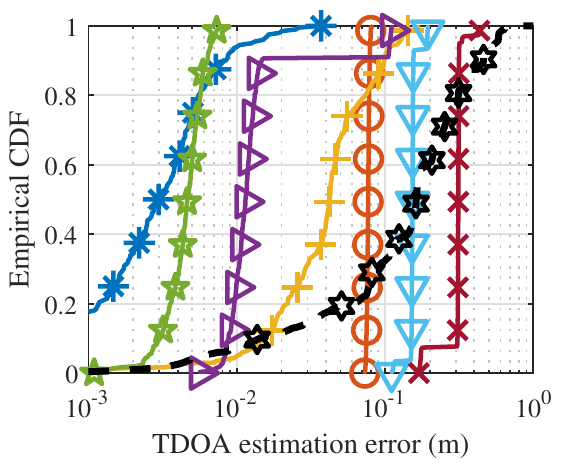}}\\
  \subfloat{\includegraphics{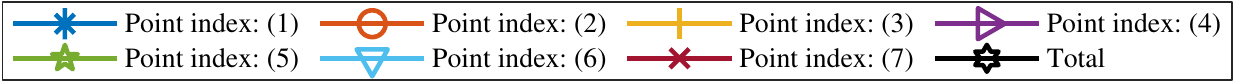}}
  \caption{Empirical CDF curves for DOA and TDOA estimation errors with data collected in the underground parking lot.
  }
  \label{fig:jade_result_indoor}
\end{figure*}

First, the JADE performance is evaluated. As stated before, errors for the differentials of the TOA estimates at these two TRPs, a.k.a the TDOA estimation errors, are calculated instead of the TOA estimation errors. The ensemble empirical CDF curves for DOA and TDOA estimation errors are shown in Fig. \ref{fig:jade_result_indoor}, along with empirical CDFs for the seven selected test points. The total \(80\)-th percentiles for DOA estimation errors at these two RRUs are \(0.88^\circ\) and \(1.49^\circ\), respectively, while the \(80\)-th percentile for TDOA estimation error is \(0.31\;\mathrm{m}\). The estimation error is highly location-dependent, which is attributed to the varied multipath effects at different locations.

\begin{figure*}[htb]
  \centering
  \subfloat[Scatter plot for positioning results of all \(300\) SRS symbols at all test points.]{\includegraphics[width=0.395\textwidth]{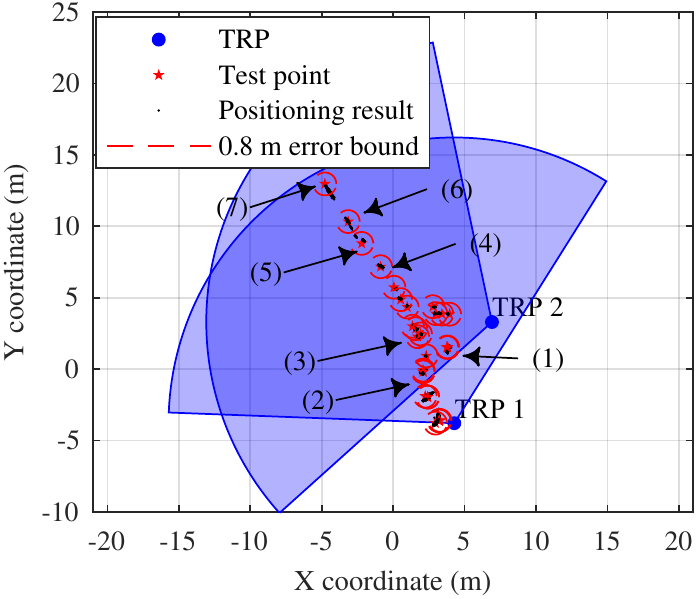}}
  \hfill
  \subfloat[Empirical CDFs for errors of triangulation positioning based on DOA estimates of two TRPs.]{\includegraphics{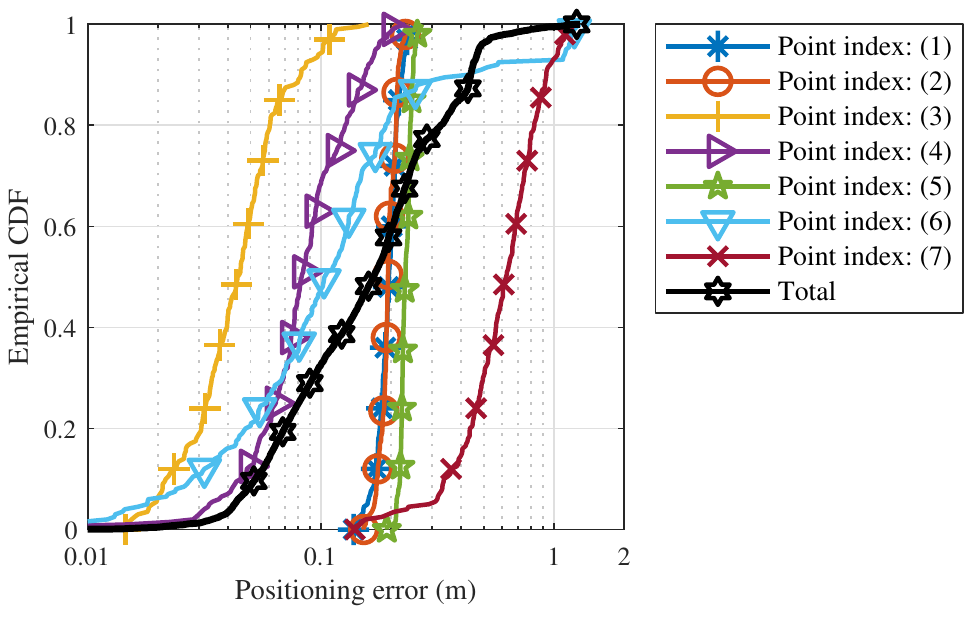}}
  \caption{Performance evaluation for triangulation positioning based on DOAs at the two established TRPs estimated by the proposed JADE method.}
  \label{fig:positioning_result_indoor}
\end{figure*}

Then the position of the UE is determined by triangulation based on the DOA estimates of these two TRPs, resulting in the positioning errors illustrated in Fig. \ref{fig:positioning_result_indoor}. Fig. \ref{fig:positioning_result_indoor}(a) indicates that the positioning results fall within the \(0.8\;\mathrm{m}\) error bounds in most cases. Similar to Fig. \ref{fig:jade_result_indoor}, Fig. \ref{fig:positioning_result_indoor}(b) depicts the empirical CDF curves of the positioning errors for the seven selected positions and the ensemble CDF curve. 
It demonstrates that the total \(90\)-th percentile of positioning error for all these \(9000\) samples is \(0.44\;\mathrm{m}\), which meets the 3GPP R17 requirement for commercial use cases (\(1\;\mathrm{m}\) for \(90\%\) cases) \cite{3gpp.38.857}.
In addition, while the current positioning results are obtained via triangulation for each individual SRS symbol, hybrid positioning methods which combine the tracking frameworks with both DOA and TDOA estimates, such as the one proposed in \cite{aernouts2020_CombiningTDoAAoA}, can be employed for further performance improvement. 



\subsection{Running Time Evaluation}
The running time of the proposed method is evaluated for the above picocell gNB and SRS configurations and is compared with that of the smoothed-MUSIC-based JADE method. The running times are all recorded by the timer of MATLAB on a PC with a \(2.60\;\mathrm{GHz}\) Intel i7 CPU and a \(16\;\mathrm{GB}\) memory. As stated at the start of Section \ref{sec:perf-eval}, the proposed pre-processing scheme with same parameters is applied to all algorithms. Therefore, their input CFR matrices have the identical dimension. The averaged running times are shown in TABLE \ref{tab:running_time}. It shows that, by using the cascading estimation scheme for JADE, the computation time is reduced by two orders of magnitude when compared with the popular smoothed-MUSIC-based JADE method. In addition, its running time can be further reduced by nearly an order of magnitude by employing FFTs for fast IAA spectrum computing.


\begin{table}[htbp]
  \caption{Running times on a PC with an Intel i7 CPU @2.60\(\;\)GHz and a 16\(\;\)GB memory}
  \begin{center}
    \begin{tabular}{cc}
      \toprule
      \bf{JADE algorithm} & \bf{Averaged running time} \\
      \midrule
      Smoothed-MUSIC-based \cite{kotaru2015_SpotFiDecimeterLeva, bazzi2016_SpatiofrequentialSm} & \(8777.0\;\mathrm{ms}\) \\
      IAA + CBF (proposed, Section \ref{sec:proposed_jade}) & \(88.1\;\mathrm{ms}\) \\
      FFT-IAA + CBF (proposed, Section \ref{sec:fastIAA}) & \(10.1\;\mathrm{ms}\) \\
      \bottomrule
    \end{tabular}
    \label{tab:running_time}    
  \end{center}
\end{table}

\section{Discussion}
\label{sec:ext}

This work proposes a JADE method in the presence of direction-dependent antenna modeling errors for picocell gNBs with ULAs established. However, the framework of the proposed method is quite adaptable to different scenarios. In this section, discussions about the extensions of the proposed framework to other potential scenarios are presented.

\subsection{Extension to Arbitrary Array Structures}
Many JADE methods require the Vandermonde array manifold and cannot be directly applied in non-uniform array configurations, such as those relied on the spatial smoothing technique \cite{kotaru2015_SpotFiDecimeterLeva, bazzi2016_SpatiofrequentialSm}, the array shift invariance property \cite{vanderveen1997_JointAngleDelaya, li2021_DecimeterLevelIndo}, or the polynomial rooting technique \cite{bazzi2015_EfficientMaximumLi}. In contrast, the proposed method is applicable for arbitrary array structures since it employs the CBF for DOA estimation. That is, it can handle the space-domain non-uniformity.

\subsection{Extension to Nonuniformly Distributed Subcarriers}
Thanks to the IAA spectral estimator, which achieves a clear and accurate spectrum even when the underlying signature structure is nonuniform \cite{Babu2010_Spectral, vu2012_NonparametricMissin}, the proposed method can also handle the frequency-domain non-uniformity. This characteristic is promising in the scenario with co-channel interferences, in which the contaminated subcarriers can be discarded and the remaining subcarriers with nonuniformly distributed pattern are exploited for TOA spectrum estimation. Additionally, in this situation, the IAA method can also be accelerated by FFTs using the approach presented in Section \ref{sec:fastIAA}. Specifically, denoting the uncontaminated CFR for \(n\)-th channel as \(\breve{\mathbf{h}}_n\) and the number of uncontaminated subcarriers as \(\breve{M}\), then \(\breve{\mathbf{h}}_n\) relates to the full CFR \(\mathbf{h}_n'\) via a selection matrix \(\mathbf{J}\in\mathbb{R}^{\breve{M}\times{}M}\) as follows:
\begin{equation}
  \breve{\mathbf{h}}_n = \mathbf{J}\mathbf{h}_n',\quad n = 1,\dots,N,
\end{equation}
where when the \(m\)-th subcarrier is not contaminated, \(\mathbf{J}\) has an element equals to \(1\) at the \(m\)-th column, otherwise its \(m\)-th column is \(\mathbf{0}_{\breve{M}}\). Similarly, the delay signature matrix for the uncontaminated CFR also relates to that for the original CFR by the same selection matrix \(\mathbf{J}\):
\begin{equation}
  \breve{\mathbf{A}}_\tau = \mathbf{J}\mathbf{A}_\tau.
\end{equation}

Accordingly, the IAA iterations of equation (\ref{eq:iaa_iter1}) and (\ref{eq:iaa_iter3}) for uncontaminated CFR can be updated to
\begin{align}
  \hat{\beta}_{n,p} &= \frac{\mathbf{a}_{\tau}^{\mathrm{H}}(\tau_p)\left[\mathbf{J}^{\mathrm{T}}\cdot\left(\breve{\mathbf{R}}_{n}^{-1}\breve{\mathbf{h}}_n\right)\right]}{\mathbf{a}_{\tau}^{\mathrm{H}}(\tau_p)\left[\mathbf{J}^{\mathrm{T}}\breve{\mathbf{R}}_{n}^{-1}\mathbf{J}\right]\mathbf{a}_{\tau}(\tau_p)}, \quad p = 1,\dots, P, \label{eq:iaa_iter1_updated}\\
    \breve{\mathbf{R}}_n &= \mathbf{J}\left[\mathbf{A}_{\tau}\hat{\mathbf{P}}_n\mathbf{A}_{\tau}^{\mathrm{H}}\right]\mathbf{J}^{\mathrm{T}},\label{eq:iaa_iter3_updated}
\end{align}
where \(\mathbf{A}_\tau\)-related operations can still be implemented by FFTs according to Section \ref{sec:fastIAA}, and \(\mathbf{J}\)-related operations can be implemented by matrix selection and matrix zero-padding, which have little impact on the computational complexity.

Therefore, the proposed method implemented with FFT-IAA can be extended effortlessly to the CFR model with nonuniformly distributed subcarriers.  


\subsection{Extension to 3-D Localization Parameter Estimation}

This paper emphasizes on the estimation of the propagation delay \(\tau\) and azimuth angle \(\theta\) of the arrived signal for 2-D positioning based on a linear array. In fact, since the proposed method separates the temporal and spatial processing, it can also be extended to estimate the delay \(\tau\), azimuth angle \(\theta\), and elevation angle \(\phi\) for positioning in a three-dimensional (3-D) space based on an arbitrary planar array. Concretely, in this situation, the 1-D CBF search expressed by equation (\ref{eq:target_fun_DOA}) is replaced with the following 2-D CBF search to determine the optimal \(\theta\) and \(\phi\) of the LOS path:
\begin{equation}
  \left(\hat{\theta}_{\mathrm{LOS}}, \hat{\phi}_{\mathrm{LOS}}\right) = \arg\max_{\left(\theta,\phi\right)}\left[\hat{\mathbf{a}}_{\theta,\phi}'(\theta,\phi)\right]^{\mathrm{H}}\mathbf{b}_{\text{LOS}},
\end{equation}
where \(\hat{\mathbf{a}}_{\theta,\phi}'(\theta,\phi)\) is the estimated actual steering-vector function of the specific planar array for input pair of \((\theta,\phi)\).

\section{Conclusion}
\label{sec:conclusion}

This study has demonstrated the potential of 5G picocell gNB for indoor positioning by investigating its array modeling errors specifically and designing an efficient JADE scheme to calibrate these errors and to provide DOA and TOA estimates in real-time.
First, based on the characteristics of the array modeling errors of typical picocell gNBs, a vector-valued function for the direction-dependent antenna error is incorporated in the signal model to capture all sorts of antenna errors.
Then the antenna error function is estimated and compensated to the ideal steering-vector to derive the actual array steering-vector for DOA estimation in the proposed JADE scheme. The proposed scheme achieves largely reduced computational complexity and storage requirement by employing a cascading processing scheme with the IAA-based TOA spectral analyzer and a CBF-based DOA estimator, which fully exploits the fact that the 5G picocell gNB only has a small-scale antenna array but has a large signal bandwidth. By further proposing a dimension-reducing pre-processing method and deriving the FFT-based version of the IAA method, an averaged running time of \(10.1\;\mathrm{ms}\) is achieved for the proposed method in a typical gNB configuration, which is nearly three-orders lower than that of the MUSIC-based JADE method. Some attentions have also drawn to the adaptability of the proposed scheme, which has shown to be able to cope with both the irregularities of the antenna array and the subcarrier distribution, and be scalable to the scenario of planar-array-based 3-D positioning.

A series of experiments was designed for performance validation. First, a field test in the anechoic chamber demonstrated the effectiveness of the calibration scheme. Then numerical simulations based on simulated multipath data and 5G channel data showed the superiority of the proposed method for DOA estimation at the cost of a slightly reduced TOA estimation performance when compared with popular 2-D super-resolution JADE methods. Lastly, according to the field test in an indoor environment, the proposed method achieved a positioning error of \(0.44\;\mathrm{m}\) for \(90\%\) cases by employing a minimum gNB configuration with only two separate TRPs. In summary, we think the proposed method is a viable option for performing real-time JADE in future 5G and beyond network for the purpose of high-accuracy indoor positioning.


\ifCLASSOPTIONcaptionsoff
  \newpage
\fi

\balance
\bibliographystyle{IEEEtran}
\bibliography{paper_iaadbf}

\end{document}